\documentclass[letterpaper,11pt]{article}
\usepackage{jheppub} 
\usepackage{slashed, subcaption, mathrsfs, braket, mathtools, float, soul}
\usepackage{bm, bbm}
\usepackage{tikz, circuitikz}
\usetikzlibrary{arrows.meta, positioning, decorations.pathmorphing}
\usepackage[compat=1.1.0]{tikz-feynman}

\usepackage{graphicx}

\DeclareCaptionLabelFormat{parens}{(#2)} 
\newcommand{\herm}{\dagger}
\newcommand{\D}{\mathrm{d}}
\newcommand{\Tr}{\text{Tr}}

\def\beq{\begin{equation}}
\def\eeq{\end{equation}}

\newcommand{\I}{\mathrm{i}}
\newcommand{\e}{\mathrm{e}}

\newcommand{\cB}{\mathcal{B}}
\newcommand{\cM}{\mathcal{M}}
\newcommand{\cMbar}{\overline{\mathcal{M}}}
\newcommand{\mm}{(m_{\mathrm{M}})}

\newcommand{\mmbar}{\cM \text{--} \cMbar}

\title{Neutrino properties from muonium-antimuonium mixing}
\author[a]{Mitrajyoti Ghosh,}
\emailAdd{mghosh2@fsu.edu}
\author[a]{Kevin Liguori,}
\emailAdd{kliguori@fsu.edu}
\author[a,b]{Takemichi Okui,}
\emailAdd{tokui@fsu.edu}
\author[a,b]{Kohsaku Tobioka}
\emailAdd{ktobioka@fsu.edu}
\affiliation[a]{Department of Physics, Florida State University, 77 Chieftan Way, Tallahassee, FL 32306-4350, USA}
\affiliation[b]{Theory Center, High Energy Accelerator Research Organization (KEK), 1-1 Oho, Tsukuba, Ibaraki 305-0801, Japan}

\abstract{The nature of neutrino mass---whether neutrinos are Dirac or Majorana particles---remains one of the central open questions in particle physics. While observation of neutrinoless double beta decay would confirm the Majorana case, the absence of a signal offers no definitive insight. In light of this, we investigate muonium–antimuonium mixing---proposed for further study at the MACE experiment---as an alternative probe of neutrino properties. We compute the mixing amplitude in the Standard Model minimally extended to include massive Dirac or Majorana neutrinos, and correct previous calculations by properly treating the relevant infrared scales. As the GIM mechanism strongly suppresses the Dirac contribution, we explore whether relaxing unitarity of the PMNS matrix can enhance the mixing without obscuring neutrino properties. Surprisingly, the answer to this question is negative. We also examine the pseudo-Dirac case---predominantly Dirac neutrinos with small Majorana masses---and find that this scenario can significantly enhance the mixing compared to the pure Dirac case, especially for normal mass ordering.} 

\begin{document}
\maketitle
\flushbottom

\section{Introduction}

One of the most important unanswered questions in fundamental physics is whether neutrinos are lepton number violating (Majorana) or conserving (Dirac) particles. 
The most popular experimental search that attempts to answer this question is neutrino-less double beta decay ($0\nu\beta\beta$)~\cite{KamLAND-Zen:2022tow,GERDA:2020xhi}. 
However, if the neutrinos are Dirac, $0\nu\beta\beta$ will not provide us with a \emph{positive} confirmation but only the absence of Majorana signals. 
For an experimental observable that can positively distinguish between Dirac vs Majorana, 
we need an observable that conserves total lepton number ($\Delta L = 0$). In addition, to further access neutrino properties not elucidated by $0\nu\beta\beta$, it would be desirable to violate lepton number for individual lepton species (i.e., $\Delta L_{e,\mu,\tau} \neq 0$).
One candidate process is mixing between muonium ($\cM$) and its CP conjugate state antimuonium ($\cMbar$), 
where $\cM$ ($\cMbar$) is a bound state of $\mu^+$ and $e^-$ ($\mu^-$ and $e^+$).
For the purpose of studying neutrino properties, muonium provides an especially clean probe free from hadronic physics, and is also experimentally accessible due to its appreciably long lifetime, which is the muon lifetime to very good approximation~\cite{ParticleDataGroup:2024cfk, Czarnecki:1999yj}. 
$\mmbar$ mixing has been studied also in other interesting contexts~\cite{Feinberg:1961zza, Glashow:1961zz, Swartz:1989qz, Hou:1995dg,Horikawa:1995ae,Willmann:1998gd, Endo:2020mev, Conlin:2020veq, Han:2021nod, Fukuyama:2021iyw, Petrov:2022wau}.

The Muonium Antimuonium Conversion Spectrometer (MACS)~\cite{Willmann:1998gd} experiment at the Paul Scherrer Institute (PSI) searched for $\mmbar$ conversion
(for earlier searches see~\cite{Amato:1968xyq, Chang:1989uk, Chatterjee:1992yi,  Huber:1988gu, Matthias:1991fw}).
Since no $\mmbar$ conversion was seen at MACS with $\sim 10^{11}$ muonium decays collected, the experiment placed an upper bound on the conversion probability $< 8.3 \times 10^{-11}$.
The proposed Muonium-to-Antimuonium Conversion Experiment (MACE)~\cite{Bai:2024skk} aims to gather $10^{14}$ muonium decays to probe muonium conversion probabilities down to $10^{-13}$.

\begin{figure}
    \begin{subfigure}[b]{0.45\textwidth}
           \begin{tikzpicture}
        \begin{feynman}
            \vertex (i1) at (-3, 1) {\(\mu^+\)};
            \vertex (i2) at (-3,-1) {\(e^-\)};
            \vertex (o1) at (3, 1) {\(e^+\)};
            \vertex (o2) at (3,-1) {\(\mu^-\)};

            \vertex (v1) at (-1.5, 1);
            \vertex (v2) at (1.5, 1);
            \vertex (v3) at (-1.5,-1);
            \vertex (v4) at (1.5,-1);

            \vertex (m1) at (0,1);
            \vertex (m2) at (0,-1);

            \diagram* {
                (i1) -- [anti fermion] (v1) --[anti fermion] (v2) -- [ anti fermion] (o1),
                (i2) -- [fermion] (v3) -- [fermion] (v4) -- [fermion] (o2),
            };

            \draw[decorate, decoration={snake, amplitude=0.15cm, segment length=0.35cm}] (v1) -- (v3);
            \draw[decorate, decoration={snake, amplitude=0.15cm, segment length=0.35cm}] (v2) -- (v4);

            \node[above] at (m1) {\(\nu_i\)};
            \node[below] at (m2) {\(\nu_j\)};

        \end{feynman}
    \end{tikzpicture}
       \caption{A Dirac type diagram}
       \label{fig:mosc_dirac}
    \end{subfigure}
    \hfill
    \begin{subfigure}[b]{0.45\textwidth}
           \begin{tikzpicture}
        \begin{feynman}
            \vertex (i1) at (-3, 1) {\(\mu^+\)};
            \vertex (i2) at (-3,-1) {\(e^-\)};
            \vertex (o1) at (3, 1) {\(\mu^-\)};
            \vertex (o2) at (3,-1) {\(e^+\)};

            \vertex (v1) at (-1.5, 1);
            \vertex (v2) at (1.5, 1);
            \vertex (v3) at (-1.5,-1);
            \vertex (v4) at (1.5,-1);

            \vertex (m1) at (0,1);
            \vertex (m2) at (0,-1);

            \diagram* {
                (i1) -- [anti fermion] (v1) --[anti fermion] (m1) -- [ fermion] (v2)--[fermion](o1),
                (i2) -- [fermion] (v3) -- [fermion](m2)--[anti fermion](v4) -- [anti fermion] (o2),

            };
            \draw[decorate, decoration={snake, amplitude=0.15cm, segment length=0.35cm}] (v1) -- (v3);
            \draw[decorate, decoration={snake, amplitude=0.15cm, segment length=0.35cm}] (v2) -- (v4);

            \draw[thick] (m1) --++ (-0.10,0.10) --++ (0.2,-0.2);
            \draw[thick] (m1) --++ (-0.10,-0.10) --++ (0.2,0.2);

            \draw[thick] (m2) --++ (-0.10,0.10) --++ (0.2,-0.2);
            \draw[thick] (m2) --++ (-0.1,-0.1) --++ (0.2,0.2);

            \node[above] at (m1) {\(\nu_i\)};
            \node[below] at (m2) {\(\nu_j\)};

        \end{feynman}
    \end{tikzpicture}
       \caption{A Majorana type diagram}
       \label{fig:mosc_maj}
    \end{subfigure}
    \caption{Examples of the Feynman diagrams that contribute to $\mmbar$ mixing via neutrinos. Arrows represent the flow of lepton number. On the left is a ``Dirac type'' diagram where lepton number is conserved on each fermion line, while on the right is a ``Majorana type'' diagram where lepton number is violated on each fermion line by the Majorana mass. }
    \label{fig:muonium_osc}
\end{figure}

In the Standard Model (SM) minimally extended with neutrino masses, $\mmbar$ mixing occurs via one or both types of box diagrams shown in Fig~\ref{fig:muonium_osc}. 
The diagram in Fig.~\ref{fig:mosc_dirac} is what we call a ``Dirac type'' diagram, where lepton arrows are preserved. 
Fig.~\ref{fig:mosc_maj} is what we refer to as a ``Majorana type'' diagram, where the lepton arrows are not preserved. 
In both diagrams, the total lepton number is conserved, $\Delta L = 0$. 
However, the muon and electron numbers are not individually conserved, $\Delta L_{e, \mu} \neq 0$. 
If neutrinos are Dirac, only the Dirac type diagrams contribute to the mixing, whereas for Majorana neutrinos both Dirac and Majorana type diagrams contribute. 

A glance at Fig.~\ref{fig:muonium_osc} tells us that the Majorana type diagram goes as $G_F^2 \sum_{i,j} (U^*_{e j})^2 U_{\mu i}^2 m_j m_i$ while the Dirac type goes as $G_F^2 (\sum_{i} U^*_{e i} U_{\mu i} m_{i}^2)^2 /m_{\mu}^2$, 
where $U_{\alpha i}$ ($\alpha = e, \mu, \tau$, $i=1,2,3$) denote components of the PMNS matrix ($U_\mathrm{PMNS}$).
Note that for the Dirac type diagrams the amplitude would vanish by the unitarity of $U_\mathrm{PMNS}$ if we did not pick up $m_i^2$. 
A calculation of the mixing amplitude exists in the literature \cite{Clark:2003tv, Fukuyama:2021iyw} for Majorana neutrinos; however, these previous calculations incorrectly handled IR scales and hence are parametrically incorrect. We will address this issue. 

The relative smallness of the Dirac type mixing amplitude compared to the Majorana type is a direct consequence of the unitarity of $U_\mathrm{PMNS}$. 
Since we do not expect the unitarity of $U_\mathrm{PMNS}$ to be an exact property in the presence of new physics, there could be contributions without any powers of neutrino mass in the Dirac amplitude. 
It is then natural to ask if relaxing this rigid requirement of a unitary PMNS matrix could lead to an enhancement in the $\mmbar$ conversion probability. 
Further, Dirac neutrinos may well have small Majorana components (i.e., pseudo-Dirac).
In this case, we expect the amplitude to go as the square of the small Majorana components, which could be larger than the above Dirac estimate. 
We will investigate these interesting and important questions. 

We organize the rest of the article as follows: In Sec.~\ref{sec:review}, we discuss $\mmbar$ conversion from a macroscopic viewpoint without considering the internal structure of muonium. 
In particular, we identify appropriate amplitudes necessary to describe $\mmbar$ mixing. 
In Sec.~\ref{sec:SM}, we calculate those amplitudes from the Dirac type and Majorana type box diagrams. 
In Sec.~\ref{sec:conversionprobs}
we then apply the results to obtain the conversion probabilities for Majorana, Dirac and pseudo-Dirac. 
We also address the question of whether relaxing the unitarity of $U_\mathrm{PMNS}$ leads to an enhancement in the rate for Dirac neutrinos. 
We make concluding remarks in Sec.~\ref{sec:conc}. 
In the appendix we provide a modern treatment of interpolating a bound state using the fundamental fields of the SM.

\section{Muonium-antimuonium mixing formalism}
\label{sec:review}
In this section, we concern ourselves with an effective description of the production, propagation, and detection of (anti)muonium without going into microscopic details. 
We begin by representing the creation of an $\cM$ at a muonium source $S$, 
and its subsequent propagation and detection as an $\cM$ or $\cMbar$ at a detector $D$ or $\bar{D}$, respectively.
That is, we would like to study the following effective quadratic lagrangian:
\beq
\mathcal{L}
= \dot{\phi}^\dag \dot{\phi} 
 - \tilde{M}^2 \phi^\dag \phi
 - \frac{\Sigma_{21}}{2} \phi \phi - \frac{\Sigma_{12}}{2} \phi^\dag \phi^\dag
 + S\phi^\dag + D \phi + \bar{D} \phi^\dag
\,,\label{eq:lagrangian:phi-phidagger}
\eeq
where $\phi$ is the muonium field with $\tilde{M} \equiv M - \I\Gamma/2$ encapsulating the muonium mass $M$ and width $\Gamma$ 
that we would have in the absence of $\cM$--$\cMbar$ mixing.
We work in the muonium rest frame and suppress the dependence on the muonium polarization. This is possible because the $\mmbar$ amplitude is independent of the polarization in the rest frame, as we will discuss later.
$\cM$--$\cMbar$ mixing is described by $\Sigma_{21}$ and $\Sigma_{12}$, 
where the former is the 1PI self-energy with an ingoing $\cM$ and an outgoing $\cMbar$,    
while the latter is that with an ingoing $\cMbar$ and an outgoing $\cM$.

The bulk of this paper will be devoted to calculating $\Sigma_{21,12}$ from the underlying microscopic physics. 
It will be found that $\Sigma_{21,12}$ can be written as  
$\Sigma_{21} = \tilde{\mu}^2 \e^{\I\alpha}$ and $\Sigma_{12} = \tilde{\mu}^2 \e^{-\I\alpha}$ 
with a common complex quantity $\tilde{\mu}^2$ and a relative phase $\e^{\pm\I\alpha}$.
The relative phase is unphysical 
as it can be completely removed by redefining $\phi \rightarrow e^{\I \alpha} \phi$, 
and similarly rephasing $S$, $D$, and $\bar{D}$.
Then, with $\Sigma_{21} = \Sigma_{12} = \tilde{\mu}^2$, 
the mixing terms in~(\ref{eq:lagrangian:phi-phidagger}) simply become $-\tilde{\mu}^2(\phi\phi + \phi^\dag \phi^\dag)/2$, 
which can be diagonalized in terms of two real fields $\phi_+$ and $\phi_-$ 
defined via $\phi = (\phi_+ + \I\phi_-) / \sqrt2$.
The lagrangian~(\ref{eq:lagrangian:phi-phidagger}) now takes the form
\beq
\mathcal{L}
= \sum_{i=+,-} \frac12 \bigl( \dot{\phi_i}^2 - \tilde{M}^2_i \phi_i^2 \bigr) 
  + \frac{1}{\sqrt2} \Bigl[ S(\phi_+ - \I\phi_-) + D(\phi_+ + \I\phi_-) + \bar{D}(\phi_+ - \I\phi_-) \Bigr]
\,,\label{eq:lagrangian:phi+phi-}
\eeq
where $\tilde{M}^2_\pm \equiv \tilde{M}^2 \pm \tilde{\mu}^2$.

We can now read off from~(\ref{eq:lagrangian:phi+phi-}) the amplitude for an $\cM$ created at the source $S$ to be detected as an $\cMbar$ at $\bar{D}$.
At $S$, the $\cM$ is either a $\phi_+$ or a $\phi_-$ with an amplitude $1/\sqrt{2}$ or $-\I / \sqrt2$, respectively.
At $\bar{D}$, the $\phi_+$ ($\phi_-$) is detected as an $\cMbar$ with an amplitude $1/\sqrt{2}$ ($-\I / \sqrt2$).
Thus, the net amplitude for $\cM \to \cMbar$ is given by
\beq
\begin{aligned}
\mathcal{A}(\cM \to \cMbar) 
&\;\propto\; \frac{1}{\sqrt2} \, \frac{1}{E^2 - \tilde{M}^2_+} \, \frac{1}{\sqrt2} 
             + \frac{-\I}{\sqrt2} \, \frac{1}{E^2 - \tilde{M}^2_-} \, \frac{-\I}{\sqrt2}  \\
&\;=\; \frac{\tilde{M}^2_+ -  \tilde{M}^2_-}{2(E^2 - \tilde{M}^2_+) (E^2 - \tilde{M}^2_-)}
 \;=\; \frac{\tilde{\mu}^2}{(E^2 - \tilde{M}^2)^2 }
\,,\label{eq:amp:MtoMbar}
\end{aligned}
\eeq
where $E$ is the energy transfer from the source to the $\cM$ in the $\cM$ rest frame.
In the last step above, we have ignored higher order $\tilde{\mu}^2$ dependence by 
assuming $|\tilde{\mu}|^2 \ll |\tilde{M}|^2$, which is an excellent approximation as $\cM$--$\cMbar$ mixing is necessarily extremely rare.
Similarly, the amplitude for an $\cM$ created at $S$ to be detected as an $\cM$ at $D$ is given by
\beq
\begin{aligned}
\mathcal{A}(\cM \to \cM) 
&\;\propto\; \frac{1}{\sqrt2} \, \frac{1}{E^2 - \tilde{M}^2_+} \, \frac{1}{\sqrt2} 
            + \frac{-\I}{\sqrt2} \, \frac{1}{E^2 - \tilde{M}^2_-} \, \frac{+\I}{\sqrt2}  \\ 
&\;=\; \frac{2 E^2 - \tilde{M}^2_+ - \tilde{M}^2_-}{2(E^2 - \tilde{M}^2_+) (E^2 - \tilde{M}^2_-)}
 \;=\; \frac{1}{E^2 - \tilde{M}^2}
\,,\label{eq:amp:MtoM}
\end{aligned}
\eeq
where the implicit proportionality factor is the same as that for $\mathcal{A}(\cM \to \cMbar)$.
In the last expression in~(\ref{eq:amp:MtoM}) we have dropped $\tilde{\mu}^2$ dependence 
as we will essentially be taking the ratio $\mathcal{A}(\cM \to \cMbar) / \mathcal{A}(\cM \to \cM)$
where the numerator is already $\mathcal{O}(\tilde{\mu}^2)$.  

From the amplitudes~(\ref{eq:amp:MtoMbar}) and~(\ref{eq:amp:MtoM}), 
we can obtain the ratio of the probability of $\cM \to \cMbar$ to that of $\cM \to \cM$, 
by taking the absolute square of each of the amplitudes~(\ref{eq:amp:MtoMbar}) and~(\ref{eq:amp:MtoM}), 
integrating each over the energy transfer $E$, and then taking the ratio:
\beq
P(\cM \to \cMbar) 
\;\equiv\;
\frac{\int\!\D E\, |\mathcal{A}(\cM \to \cMbar)|^2}{\int\!\D E\, |\mathcal{A}(\cM \to \cM)|^2}
\;=\; \frac{|\tilde{\mu}^2|^2}{2M^2 \Gamma^2}
\;=\; \frac{\bigl| \Sigma_{12} \Sigma_{21} \bigr|}{2M^2 \Gamma^2}
\,.\label{eq:prob}
\eeq
This ratio is often referred to as the ``conversion probability'' in the literature.
It is also common in the literature to split $\tilde{\mu}^2$ into two parts 
as $\tilde{\mu}^2 = \tilde{M}(\Delta M - \I\Delta\Gamma/2)$.
This splitting would be useful to make if the $\cM$--$\cMbar$ oscillation frequency $\Delta M$ were a good observable. This is not the case, however, as $(\Delta M)^{-1}$ is necessarily much longer than the muon lifetime.

In the following sections we will calculate the self-energy $\Sigma_{21}$ (and similarly $\Sigma_{12}$) by breaking it up into three parts: 
\beq 
-\I\Sigma_{21}^{\mu \nu} = \hat{\Gamma}^{\mu}_{d \dot{c}} \left(\I\cB^{ \dot{c}d \dot{a}b } \right)\hat{\Gamma}^{\nu}_{b \dot{a} }
\,,\label{eq:combine}
\eeq
where $\dot{a}, \dot{c}$ ($b,d$) are right-handed (left-handed) Weyl spinor indices. 
The meaning of this decomposition is most easily explained diagrammatically as in Fig.~\ref{fig:Sigma}. 
The $\I\cB$ represents box diagrams like those depicted in Fig.~\ref{fig:muonium_osc}, while the $\hat{\Gamma}^\nu$ interpolates muonium to $e^-$ and $\mu^+$. That is, the $\hat{\Gamma}^\nu$ module captures the physics of the $\mu^+e^-$ interactions which form the bound muonium state. Further, the LSZ reduction has been applied to amputate the muonium propagator from $\hat{\Gamma}^\nu$. 
To calculate $\I\cB$ and $\hat{\Gamma}^\nu$ in the rest frame of (anti)muonium, 
we will ignore corrections of ${\cal O}(m_e/m_{\mu})$ or smaller. 
In this approximation, the muonium mass and width are given by $M = m_{\mu}$ and $\Gamma = \Gamma_\mu = G_F^2 m_\mu^5/192\pi^3$. 
Also, the muonium 4-momentum is approximated by $P = (m_{\mu},0,0,0)$, while 
the 4-momentum of the $\mu^+$ going into the $\I\cB$ module and that of the $\mu^-$ going out of $\I\cB$ are both given by $P$, 
and the ingoing $e^-$'s and outgoing $e^+$'s 4-momenta by $(0,0,0,0)$.
Thus, the only external scale that goes into the loop calculation is $m_\mu$.
We work in these approximations throughout the paper.

The $\hat{\Gamma}$ module is purely QED and has nothing to do with neutrinos. 
Its computation is described in the appendix, and the result is
\begin{equation}
    \begin{aligned}
        \hat{\Gamma}^{\mu}_{b \dot{a}}(P) = - \frac{\left| \psi(0) \right|}{2 m_\mu^{3/2}} (P \!\cdot\! \sigma \; \bar{\sigma}^{\mu}  P \!\cdot\! \sigma)_{b \dot{a}}^{\phantom\dag} \, \, , \label{eq:gamma_nu}
    \end{aligned}
\end{equation}
where $\psi(0)$ is the ground-state wave function of the $e^-$ evaluated at the position of the $\mu^+$.
Note that the ground state is quadruply degenerate as it can be in a spin singlet or triplet. 
This degeneracy is labelled by the 4-vector indices on $\hat{\Gamma}^\mu$ and $\Sigma_{12, 21}^{\mu\nu}$. 
We will see later that in the rest frame of (anti)muonium we have $\Sigma_{12, 21}^{\mu\nu} \propto g^{\mu\nu}$.
Since all polarizations have the same $\cal M$--$\bar{\cal M}$ amplitude in the muonium rest frame, we have only considered the scalar muonium case in the lagrangians ~(\ref{eq:lagrangian:phi-phidagger}) and (\ref{eq:lagrangian:phi+phi-}).

\begin{figure}[t]
    \begin{subfigure}[t]{0.45\textwidth}
               \begin{tikzpicture}

            \node at (-5.5, 0) { \huge $-\I\Sigma_{21}^{\mu\nu} (P^2)=$};

            \draw[->] (-2.8, 0) -- (-2, 0);
            \node at (-2.4, 0.2) {\small $P$};

            \draw (-1,1) arc[start angle=90, end angle=270, radius=1cm] -- (-1,-1)--(-1,1);

            \draw (0,1) -- (2,1) -- (2,-1) -- (0,-1) -- cycle;

            \draw (3,1) arc[start angle=90, end angle=-90, radius=1cm] -- (3,-1)--(3,1);
    
            \draw[->] (4, 0) -- (4.8, 0);
            \node at (4.4, 0.2) {\small $P$};

            \begin{feynman}
                \vertex (i1) at (-1, 0.7);
                \vertex (i2) at (-1,-0.7) ;
                \vertex (o1) at (3, 0.7) ;
                \vertex (o2) at (3,-0.7) ;
                \vertex (v1) at (0, 0.7);
                \vertex (v2) at (0,-0.7) ;
                \vertex (v3) at (2, 0.7) ;
                \vertex (v4) at (2,-0.7) ;

                \diagram* {
                    (i1) -- [anti fermion] (v1) ,
                    (i2) -- [fermion] (v2) ,
                    (v4) -- [anti fermion] (o2),
                    (v3) -- [fermion] (o1) ,
                
                };

            \end{feynman}

            \node at (-1.5, 0) {\huge $\hat{\Gamma}^{\nu}$};
            \node at (1, 0) {\huge $\I\cB$};
            \node at (+3.5, 00) {\huge $\hat{\Gamma}^{\mu}$};

            \node at (-1.2, 0.7) {$\dot{a}$};
            \node at (-1.2, -0.7) {$b$};
            \node at (0.2, 0.7) {$\dot{a}$};
            \node at (0.2, -0.7) { $b$};
            \node at (1.8, 0.7) { $\dot{c}$};
            \node at (1.8, -0.7) { $d$};
            \node at (3.2, 0.7) { $\dot{c}$};
            \node at (3.2, -0.7) { $d$};
            \node at (-0.5, -1) { $e^{-}$};
            \node at (-0.5, 1) { $\mu^{+}$};
            \node at (2.5, 1) { $\mu^{-}$};
            \node at (2.5, -1) { $e^+$};

    \end{tikzpicture}  
    \end{subfigure}
       \vskip 0em
    \begin{subfigure}[b]{0.45\textwidth}
        \tikzset{
    Vmu/.style={
        dot,
        scale=#1, 
        label={[shift={(-0.2cm, 0.0cm)}]above:$V^{\nu}$},
    }
}

\begin{tikzpicture}

    \draw[->] (-4, 0) -- (-3, 0);
    \node at (-3.5, 0.25) {\small $P$};

    \draw (-2,1) arc[start angle=90, end angle=270, radius=1cm] -- (-2,-1) -- (-2,1);

     \node at (0, 0) {$=$};

    \draw[->] (0.9, 0) -- (1.9, 0);
    \node at (1.5, -0.25) {\small $P$};

    \begin{feynman}
        \vertex (l1) at (-2, 0.75);
        \vertex (l2) at (-2, -0.75);
        \vertex (l3) at (-0.5, 0.75);
        \vertex (l4) at (-0.5, -0.75);
        \vertex[dot, scale=1, label={[shift={(-0.2cm, 0.0cm)}]above:$V^{\nu}$}] (r0) at (2, 0) {};
        \vertex (r1) at (3, 1);
        \vertex (r2) at (3, -1);
        \vertex (r3) at (4, 1);
        \vertex (r4) at (4, -1);
        \vertex (r5) at (6, 1);
        \vertex (r6) at (6, -1);
        \vertex (r7) at (7, 1);
        \vertex (r8) at (7, -1);

        \diagram* {
            (l1) -- [anti fermion] (l3),
            (l2) -- [fermion] (l4),
            (r0) -- [quarter left, anti fermion] (r1) -- [anti fermion] (r3) -- [anti fermion] (r5) -- [anti fermion] (r7),
            (r0) -- [quarter right, fermion] (r2) -- [fermion] (r4) -- [fermion] (r6) -- [fermion] (r8),
        };
    \end{feynman}

    \draw[decorate, decoration={snake, amplitude=0.1cm, segment length=0.3cm}] (r1) -- (r2);
    \draw[decorate, decoration={snake, amplitude=0.1cm, segment length=0.3cm}] (r3) -- (r4);
    \draw[decorate, decoration={snake, amplitude=0.1cm, segment length=0.3cm}] (r5) -- (r6);
    
    \node at (-2.5, 0) {\huge $\hat{\Gamma}^{\nu}$};
    \node at (5, 0) {\huge $\cdots$};

    \node at (-2.25, 0.75) {$\dot{a}$};
    \node at (-2.25, -0.75) {$b$};
    \node at (-1.25, 1) {$\mu^+$};
    \node at (-1.25, -1) {$e^-$};
    \node at (7.25, 1) {$\dot{a}$};
    \node at (7.25, -1) { $b$};
    \node at (5, 1.25) {$\mu^+$};
    \node at (5, -1.25) {$e^-$};

    \draw (7.5, 1.25) -- (7.5, -1.25);
    \node at (8, -1) {\text{amp.}};

\end{tikzpicture}    
    \label{fig:GammaHat}
    \end{subfigure}
    \caption{\label{fig:Sigma} 
    The top diagram shows schematically the three parts of $-\I\Sigma_{21}^{\mu \nu}(P^2)$ as defined in~(\ref{eq:combine}). 
    The sum of box diagrams like those depicted in Fig.~\ref{fig:muonium_osc}
    is represented by $\I\cB$,   
    while $\hat{\Gamma}^\nu$ interpolates a muonium of 4-momentum $P$ to a $\mu^+$ and an $e^-$, as shown in the bottom diagram. 
    The vertex is given by $V^\nu = \gamma^\nu P_L$, and ``amp'' indicates that
    the muonium propagator has been already amputated from $\hat{\Gamma}$ as per the LSZ reduction.
    The internal $\mu$ and $e$ propagators are included in $\hat{\Gamma}$, not in $\I\cB$.
    }
\end{figure}

Before closing this section, we would like to make an important conceptual remark on the validity of the expression~(\ref{eq:prob}). 
As it is clear from the derivation above, the expression~(\ref{eq:prob}) assumes that the $\cM$ ($\cMbar$) is detected literally as an $\cM$ ($\cMbar$) at $D$ ($\bar{D}$).
However, this is not the case in $\cM$--$\cMbar$ conversion experiments like MACS~\cite{Willmann:1998gd} and the proposed MACE~\cite{Bai:2024skk}.
In those experiments, 
an $\cM$ ($\cMbar$) is detected via its \emph{decay products} $f$ ($\bar{f}$). Specifically, the detection relies on the expectation that the decay of $\cM$ at rest produces an ultrarelativistic $e^+$ from $\mu^+$ decay and a slow $e^-$.
When $\cMbar$ decays, the expectation for the energies of the $e^\pm$ are reversed.

Relying on the kinematics to differentiate between $\cM$ and $\cMbar$ decays suffers from an inherent background. The expected kinematics for $f$ can be mimicked by $\bar{f}$, or vice versa, if the slow $e^\pm$ undergoes hard photon exchange. Still, $f$ and $\bar{f}$ appear to be distinct final states as $f$ and $\bar{f}$ should contain $\bar{\nu}_\mu \nu_e$ and $\nu_\mu \bar{\nu}_e$, respectively. However, the neutrinos can change flavors or, if they are Majorana, can lose distinction between $\bar{\nu}$ and $\nu$, thereby causing $\bar{\nu}_{\mu} \nu_e$ and $\nu_\mu \bar{\nu}_e$ to mix. Therefore, the amplitude for the background $\cM \to f \to \bar{f}$ must be added to the signal amplitude $\cM \to \cMbar \to \bar{f}$. Then, the question is whether the total amplitude $\cM \to \bar{f}$ dominated by the signal or the background. We address this important question in a separate paper~\cite{paper-in-preparation}.
 
For the purpose of the present paper, we simply assume that we identify $\cM$ versus $\cMbar$ before they decay, e.g., by ionization into unbound $\mu^+$ and $e^-$ versus unbound $\mu^-$ and $e^+$, which are clearly distinct final states and hence do not interfere.

\section{Dirac and Majorana type box diagrams in $\mmbar$ mixing}
\label{sec:SM}
In this section, we present our calculation of the two types of box diagrams---the Dirac type and Majorana type---that contribute to $\cM$--$\cMbar$ mixing.
In a Dirac type diagram, lepton number is conserved on \emph{each} lepton line, 
while in a Majorana type diagram it is violated by $\pm 2$ units on each lepton line although the diagram as a whole conserves lepton number.  
The Dirac and Majorana type diagrams are shown in Figs.~\ref{fig:dirac_diag} and~\ref{fig:majorana_diag}, 
and computed in sections~\ref{sec:DiracDiagrams} and~\ref{sec:MajoranaDiagrams}, respectively. 
If the neutrinos are Dirac fermions, only the Dirac type diagrams contribute to $\cM$--$\cMbar$ mixing as Dirac mass by definition conserves lepton number on each fermion line.
If the neutrinos are Majorana fermions, both Dirac and Majorana diagrams contribute to the mixing.
The combined mixing amplitudes and the final $\cM$--$\cMbar$ conversion probabilities will be presented in section~\ref{sec:conversionprobs}.

\subsection{Dirac type diagrams}
\label{sec:DiracDiagrams}
The amplitude of the Dirac type diagrams from Fig.~\ref{fig:dirac_diag} is given, up to corrections of ${\cal O} (m_\mu^2/m_W^2)$, by
\beq 
\I\cB_\mathrm{D} ^{\dot{c}d \dot{a}b  } 
= -32 G_F ^2 \, (m^2_{e\mu})^2 \, {\cal I}_\mathrm{D}^{\mu\nu} 
\!\left(  \bar{\sigma}_\mu^{\dot{a}d} \bar{\sigma}_\nu^{\dot{c}b} -  \bar{\sigma}_\mu^{\dot{a}b} \bar{\sigma}_\nu^{\dot{c}d} \right)\!
\,,\label{eq:diracbox}
\eeq
where 
\beq
m^2_{e\mu} \equiv \sum _{i} U_{e i}^* \, U_{{\mu} i}^{\phantom*} \, m_i^2
\,.\label{eq:m2emu}
\eeq
\begin{figure}[t]
    \centering
    \begin{subfigure}[b]{0.45\textwidth}
           \begin{tikzpicture}
        \begin{feynman}
            \vertex (i1) at (-3, 1) {\(\mu^+\)};
            \vertex (i2) at (-3,-1) {\(e^-\)};
            \vertex (o1) at (3, 1) {\(e^+\)};
            \vertex (o2) at (3,-1) {\(\mu^-\)};

            \vertex (v1) at (-1.5, 1);
            \vertex (v2) at (1.5, 1);
            \vertex (v3) at (-1.5,-1);
            \vertex (v4) at (1.5,-1);

            \vertex (l1) at (-1.75, 1);
            \vertex (l2) at (1.75, 1);
            \vertex (l3) at (-1.75,-1);
            \vertex (l4) at (1.75,-1);
            
            \vertex (m1) at (0,1);
            \vertex (m2) at (0,-1);

            \diagram* {
                (i1) -- [anti fermion] (v1) --[anti fermion] (v2) -- [ anti fermion] (o1),
                (i2) -- [fermion] (v3) -- [fermion] (v4) -- [fermion] (o2),
            };

            \draw[decorate, decoration={snake, amplitude=0.15cm, segment length=0.35cm}] (v1) -- (v3);
            \draw[decorate, decoration={snake, amplitude=0.15cm, segment length=0.35cm}] (v2) -- (v4);

            \node[above] at (m1) {\(\nu_i\)};
            \node[below] at (m2) {\(\nu_j\)};
            \node[above] at (l1) {$\dot{a}$};
            \node[above] at (l2) {$d$};
            \node[below] at (l3) {$b$};
            \node[below] at (l4) {$\dot{c}$};

        \end{feynman}
    \end{tikzpicture}
    \end{subfigure}
    \hfill
    \begin{subfigure}[b]{0.45\textwidth}
           \begin{tikzpicture}
        \begin{feynman}
            \vertex (i1) at (-3, 1) {\(\mu^+\)};
            \vertex (i2) at (-3,-1) {\(e^-\)};
            \vertex (o1) at (3, 1) {\(e^+\)};
            \vertex (o2) at (3,-1) {\(\mu^-\)};

            \vertex (v1) at (-1.5, 1);
            \vertex (v2) at (1.5, 1);
            \vertex (v3) at (-1.5,-1);
            \vertex (v4) at (1.5,-1);

            \vertex (l1) at (-1.75, 1);
            \vertex (l2) at (1.75, 1);
            \vertex (l3) at (-1.75,-1);
            \vertex (l4) at (1.75,-1);

            \vertex (m1) at (-1.5,0);
            \vertex (m2) at (1.5,0);

            \diagram* {
                (i1) -- [anti fermion] (v1) --[anti fermion] (v3) -- [ anti fermion] (i2),
                (o1) -- [fermion] (v2) -- [fermion] (v4) -- [fermion] (o2),
            };

            \draw[decorate, decoration={snake, amplitude=0.1cm, segment length=0.31cm}] (v1) -- (v2);
            \draw[decorate, decoration={snake, amplitude=0.1cm, segment length=0.31cm}] (v3) -- (v4);

            \node[left] at (m1) {\(\nu_i\)};
            \node[right] at (m2) {\(\nu_j\)};
            \node[above] at (l1) {$\dot{a}$};
            \node[above] at (l2) {$d$};
            \node[below] at (l3) {$b$};
            \node[below] at (l4) {$\dot{c}$};

        \end{feynman}
    \end{tikzpicture}
    \end{subfigure}
    \caption{\label{fig:dirac_diag}
    Dirac-type box diagrams giving rise to $\cM$--$\cMbar$ mixing. 
    Lepton number is conserved on each fermion line. 
    These diagrams contribute to the amplitude in both the Dirac and Majorana cases.}
\end{figure}
Each fermion line provides one factor of $m^2_{e\mu}$. 
Note that the amplitude would vanish by the unitarity of $U_\mathrm{PMNS}$ if we had ignored the neutrino mass in either one of the fermion lines (the GIM mechanism). 
Thus, the leading non-vanishing contribution comes from Taylor-expanding both neutrino propagators as $1/(p^2 - m_i^2) = 1/p^2 + m_i^2/(p^2)^2 + \cdots$ and picking up one factor of $m_i^2$ for each fermion line,  
which gives rise to the form~(\ref{eq:m2emu}).
The loop integral ${\cal I}_\mathrm{D}^{\mu\nu}$ is given by
\beq 
{\cal I}_\mathrm{D}^{\mu\nu} 
\equiv \int\! \frac{\D ^4 q}{(2\pi)^4} \, \frac{q^{\mu} (q + P)^{\nu} }{(q^2)^2 [ (q + P)^2]^2}  = {\I \over 32 \pi^2  P^2 } \left(g^{\mu \nu} -  {2 P^{\mu} P^{\nu} \over P^2} \right)\! 
\,,\label{eq:diracI} 
\eeq
where $m_W$ has been sent to infinity as the integral remains UV convergent without $m_W$.
The IR convergence is provided by $m_\mu$ in $P=(m_{\mu},0,0,0)$. We then find 
\beq \I\cB_\mathrm{D}^{\dot{c}d  \dot{a}b } 
= \I\frac{2 G_F^2}{\pi^2} \, \epsilon ^{\dot{c} \dot{a}} \epsilon^{db} \frac{(m^2_{e\mu})^2}{m_\mu^2}  
\,.\label{eq:boxdirac}
\eeq
%

\subsection{Majorana type diagrams}
\label{sec:MajoranaDiagrams}
The amplitude of the Majorana type diagrams from Fig.~\ref{fig:majorana_diag} is given by
\beq 
\I\cB_\mathrm{M}^{\dot{c}d \dot{a}b} 
= \frac{g^4}{4} \, m_{ee}^* m_{\mu\mu} \!
   \left[ \!\left( 4{\cal I}_{\mathrm{M}1} + {1 \over m_W^4} \, {\cal I}_{\mathrm{M}2} \right)\! 2\epsilon^{\dot{c} \dot{a}} \epsilon^{db} 
         + {4 \over m_W^2} \, {\cal I}_{\mathrm{M}3}^{\mu\nu} \!
           \left( \bar{\sigma}_{\mu}^{\dot{a}d} \bar{\sigma}_{\nu}^{\dot{c}b} 
                 - \bar{\sigma}_{\mu}^{\dot{a}b} \bar{\sigma}_{\nu}^{\dot{c}d}
           \right)\! 
   \right],
\eeq
where
\beq 
m_{\ell\ell} \equiv \sum_{i} (U_{\ell i})^2 \, m_i
\,. 
\eeq
The one factor of $m_i$ for each fermion line comes from picking up the Majorana mass in the numerator of the neutrino propagator.
Once $m_i$ is picked up, we can safely ignore the $m_i^2$ in the denominator of the propagator as the IR divergence will be cut off by $m_\mu$ in $P$. 
Then, up to terms suppressed by powers of $m_\mu/m_W$, the integrals ${\cal I}_{\mathrm{M}1,2,3}$ are given by 
\beq 
{\cal I}_{\mathrm{M}1} 
\equiv \int\! {\D^4 q \over (2\pi)^4} \, {1 \over q^2 (q + P)^2 (q^2 -m_W^2)^2 }
= \frac{\I}{16\pi^2 m_W^4} \!\left( \log\frac{m_W^2}{m_\mu^2} + \I\pi \right)\!
\,, \label{eq:IM1}
\eeq
\beq 
{\cal I}_{\mathrm{M}2} 
\equiv \int\! {\D^4 q \over (2\pi)^4} \, {(q^2)^2 \over q^2 (q + P)^2 (q^2 -m_W^2)^2 }
= \frac{\I}{16\pi^2} \log\frac{\Lambda^2}{m_W^2}
\,, \label{eq:IM2}
\eeq
\beq
{\cal I}_{\mathrm{M}3}^{\mu\nu} 
\equiv \int\! {\D^4 q \over (2\pi)^4} \, {q^{\mu} q^{\nu} \over q^2 (q + P)^2 (q^2 -m_W^2)^2 }
= \frac{\I}{16\pi^2 m_W^2} \frac{-g^{\mu\nu}}{4}
\,, \label{eq:IM3}
\eeq
where $\Lambda$ in ${\cal I}_{\mathrm{M}2}$ is the cutoff of the effective theory, that is,   
a proxy for the mass scale of the UV physics responsible for generating the Majorana neutrino masses.
(For example, if we calculate the corresponding box diagrams in a UV-complete seesaw model 
with heavy singlet Majorana fermions with a common mass scale $m_N$, 
the integral ${\cal I}_{\mathrm{M}2}$ would now become convergent with $\Lambda$ replaced by $m_N$.)
Thus, we find 
\beq
\I \cB_\mathrm{M} ^{\dot{c}d \dot{a}b } 
= \I \frac{2 G_F^2}{\pi^2} \, \epsilon^{\dot{c} \dot{a}} \epsilon^{db} \, m_{ee}^* m_{\mu\mu} 
   \!\left( 4 \log\frac{m_W}{m_\mu} + \log\frac{\Lambda}{m_W} + 2 \pi\I \right)\! 
\,,\label{eq:boxmajorana}
\eeq
where a scheme-dependent additive constant in the real part has been absorbed into $\Lambda$.

\begin{figure}[t]
    \centering
    \begin{subfigure}[b]{0.45\textwidth}
           \begin{tikzpicture}
        \begin{feynman}
            \vertex (i1) at (-3, 1) {\(\mu^+\)};
            \vertex (i2) at (-3,-1) {\(e^-\)};
            \vertex (o1) at (3, 1) {\(\mu^-\)};
            \vertex (o2) at (3,-1) {\(e^+\)};

            \vertex (v1) at (-1.5, 1);
            \vertex (v2) at (1.5, 1);
            \vertex (v3) at (-1.5,-1);
            \vertex (v4) at (1.5,-1);

            \vertex (l1) at (-1.75, 1);
            \vertex (l2) at (1.75, 1);
            \vertex (l3) at (-1.75,-1);
            \vertex (l4) at (1.75,-1);
            
            \vertex (m1) at (0,1);
            \vertex (m2) at (0,-1);

            \diagram* {
                (i1) -- [anti fermion] (v1) --[anti fermion] (m1) -- [ fermion] (v2)--[fermion](o1),
                (i2) -- [fermion] (v3) -- [fermion](m2)--[anti fermion](v4) -- [anti fermion] (o2),

            };
            \draw[decorate, decoration={snake, amplitude=0.15cm, segment length=0.35cm}] (v1) -- (v3);
            \draw[decorate, decoration={snake, amplitude=0.15cm, segment length=0.35cm}] (v2) -- (v4);

            \draw[thick] (m1) --++ (-0.10,0.10) --++ (0.2,-0.2);
            \draw[thick] (m1) --++ (-0.10,-0.10) --++ (0.2,0.2);

            \draw[thick] (m2) --++ (-0.10,0.10) --++ (0.2,-0.2);
            \draw[thick] (m2) --++ (-0.1,-0.1) --++ (0.2,0.2);

            \node[above] at (m1) {\(\nu_i\)};
            \node[below] at (m2) {\(\nu_j\)};
            \node[above] at (l1) {$\dot{a}$};
            \node[above] at (l2) {$\dot{c}$};
            \node[below] at (l3) {$b$};
            \node[below] at (l4) {$d$};

        \end{feynman}
    \end{tikzpicture}
    \end{subfigure}
    \begin{subfigure}[b]{0.45\textwidth}
           \begin{tikzpicture}
        \begin{feynman}
            \vertex (i1) at (-3, 1) {\(\mu^+\)};
            \vertex (i2) at (-3,-1) {\(e^-\)};
            \vertex (o1) at (3, 1) {\(\mu^-\)};
            \vertex (o2) at (3,-1) {\(e^+\)};

            \vertex (v1) at (-1.5, 1);
            \vertex (v2) at (1.5, 1);
            \vertex (v3) at (-1.5,-1);
            \vertex (v4) at (1.5,-1);

            \vertex (l1) at (-1.75, 1);
            \vertex (l2) at (1.75, 1);
            \vertex (l3) at (-1.75,-1);
            \vertex (l4) at (1.75,-1);
            
            \vertex (m1) at (0,1);
            \vertex (m2) at (0,-1);

            \diagram* {
                (i1) -- [anti fermion] (v1) --[anti fermion] (m1) -- [ fermion] (v2)--[fermion](o1),
                (i2) -- [fermion] (v3) -- [fermion](m2)--[anti fermion](v4) -- [anti fermion] (o2),

            };
            \draw[decorate, decoration={snake, amplitude=0.15cm, segment length=0.35cm}] (v1) -- (v4);
            \draw[decorate, decoration={snake, amplitude=0.15cm, segment length=0.35cm}] (v2) -- (v3);

             \draw[thick] (m1) --++ (-0.10,0.10) --++ (0.2,-0.2);
            \draw[thick] (m1) --++ (-0.10,-0.10) --++ (0.2,0.2);

            \draw[thick] (m2) --++ (-0.10,0.10) --++ (0.2,-0.2);
            \draw[thick] (m2) --++ (-0.1,-0.1) --++ (0.2,0.2);
            \node[above] at (m1) {\(\nu_i\)};
            \node[below] at (m2) {\(\nu_j\)};
            \node[above] at (l1) {$\dot{a}$};
            \node[above] at (l2) {$\dot{c}$};
            \node[below] at (l3) {$b$};
            \node[below] at (l4) {$d$};

        \end{feynman}
    \end{tikzpicture}
    \end{subfigure}
    \caption{\label{fig:majorana_diag}
    Majorana-type box diagrams which give rise to $\cM$--$\cMbar$ mixing.
    Lepton number is violated by $\pm 2$ units on each fermion line by the Majorana mass.
    These diagrams contribute to the amplitude only if the neutrinos are Majorana.}
\end{figure} 

Our result differs from given in~\cite{Clark:2003tv, Fukuyama:2021iyw}.
In our calculation the IR divergence in ${\cal I}_{\mathrm{M}1}$ is cut off by the scale $m_\mu$ in $P$, where $P$ necessarily enters the box diagram through the nearly on-shell, external muon line.  
The calculations in the above references ignore external 4-momenta.
Consequently, their IR divergence is cut off by $m_\nu$, resulting in much larger logarithms, $\log(m_W^2 / m^2_\nu)$.

\section{Mixing amplitudes and conversion probabilities}
\label{sec:conversionprobs}
We begin this section by computing $\cM$--$\cMbar$ conversion probabilities for Dirac and Majorana neutrinos in sections~\ref{sec:prob_dirac} and \ref{sec:prob_majorana}, respectively.
We will find that the probability is small, but especially so for Dirac neutrinos 
due to unitarity of $U_\mathrm{PMNS}$ (the GIM cancellation). 
However, before addressing the question of unitarity, 
we will first explore in section~\ref{sec:prob_pseudodirac} whether the Dirac probability can be enhanced by making the neutrinos pseudo-Dirac, i.e., by introducing small Majorana mass.
In Sec.~\ref{sec:non-unitary} we will then ask whether relaxing the unitarity of $U_\mathrm{PMNS}$ can enhance the mixing for Dirac neutrinos.

\subsection{Dirac neutrinos}
\label{sec:prob_dirac}
For Dirac neutrinos, only the Dirac type diagrams contribute. 
Thus, we have
\beq
\I\cB_{\text{Dirac}}^{\dot{c}d  \dot{a}b } = \I\cB_\mathrm{D} ^{\dot{c}d  \dot{a}b }
\,.
\eeq
Then, Eqs.~\eqref{eq:combine}, \eqref{eq:gamma_nu}, and~\eqref{eq:boxdirac} lead to 
\beq
-\I {\Sigma}_{21, \mathrm{D}}^{\mu \nu} 
= \frac{2 G_F^2}{\pi^2} \frac{(m^2_{e\mu})^2}{m_\mu^2} \I{\cal S}^{\mu \nu}
\,,
\eeq
where 
\begin{equation}
    \begin{aligned}
        \I\mathcal{S}^{\mu \nu}
        &\equiv \I \frac{\left| \psi(0) \right|^2}{4 m_\mu^3} 
                (P \!\cdot\! \sigma \, \bar{\sigma}^{\mu} P \!\cdot\! \sigma)_{d \dot{c}} \,
                \epsilon^{\dot{c} \dot{a}} \epsilon^{db} \, 
                (P \!\cdot\! \sigma \, \bar{\sigma}^{\nu} P \!\cdot\! \sigma)_{b \dot{a}} \\
        &= \I \frac{m_{\mu} \!\left| \psi(0) \right|^2}{2} g^{\mu \nu} 
        \,,\label{eq:atilde}
    \end{aligned} 
\end{equation}
which represents the IR physics associated with the $\mu^\pm e^\mp$ bound states of (anti)muonium.
For the ground state, we have $|\psi(0)|^2 = 1/(\pi a_0^3)$ with the Bohr radius $a_0 \equiv 1 / (m_e\alpha)$ 
in our approximation of neglecting corrections of order ${\cal O}(m_e/m_{\mu})$. 
Therefore, we have 
\beq
\Sigma^{\mu \nu}_{21, \mathrm{D}} 
= -\frac{m_\mu G_F^2}{\pi^3 a_0^3} \frac{(m^2_{e\mu})^2}{m_\mu^2} g^{\mu \nu}
\,,
\eeq
and $\Sigma^{\mu\nu}_{12, \mathrm{D}}$ is given by $\Sigma^{\mu\nu}_{21, \mathrm{D}}$ with the replacement $m^2_{e\mu} \to (m^2_{e\mu})^*$.
The appearance of $g^{\mu\nu}$ implies that the conversion probability~(\ref{eq:prob}) is polarization independent, 
so for each polarization we find 
\beq
P(\cM \to \cMbar)_\text{Dirac} 
= \frac{18432}{(m_\mu a_0)^6} 
\left| \frac{m^2_{e\mu}}{m_\mu^2} \right|^4
= 2.2 \times 10^{-95} \!\left( \frac{|m^2_{e\mu}|}{(10^{-1}\>\mathrm{eV})^2} \right)^{\!\!4}\! 
\,.\label{eq:pd}
\eeq
%

\subsection{Majorana neutrinos}
\label{sec:prob_majorana}
For Majorana neutrinos, all four diagrams in Figs.~\ref{fig:dirac_diag} and \ref{fig:majorana_diag} 
contribute to $\cM$--$\cMbar$ mixing. 
Thus, we have
\beq
\I\cB^{\dot{c}d  \dot{a}b }_{\text{Majorana}} 
= \I\cB_\mathrm{M}^{\dot{c}d \dot{a}b} + \I\cB_\mathrm{D}^{\dot{c}d \dot{a}b} 
\,.
\eeq
Comparing the expressions~\eqref{eq:boxdirac} and~\eqref{eq:boxmajorana} tells us that $\cB_\mathrm{D}$ is 
suppressed by a factor of ${\cal O}((m_\nu/m_\mu)^2)$ with respect to $\cB_\mathrm{M}$.
Thus, we simply neglect $\I\cB_\mathrm{D}^{\dot{c}d \dot{a}b}$ above.
 
Since the spinor structure of  $\mathcal{B}_\mathrm{M}^{\dot{c}d \dot{a}b} $ is the same as 
that of $\mathcal{B}_\mathrm{D}^{\dot{c}d \dot{a}b}$, 
we can simply repeat the steps taken in section~\ref{sec:prob_dirac} and obtain
\beq
\Sigma_{21, \mathrm{M}}^{\mu\nu} 
= -\frac{m_\mu G_F^2}{\pi^3 a_0^3} m_{ee}^* m_{\mu\mu} 
  \!\left( 4 \log\frac{m_W}{m_\mu} + \log\frac{\Lambda}{m_{W}} + 2\pi\I \right)\! g^{\mu \nu}
\,,\label{eq:Sigma21:Majo}
\eeq
and $\Sigma_{12, \mathrm{M}}^{\mu\nu}$ is given by $\Sigma_{21, \mathrm{M}}^{\mu\nu}$ 
with the replacement $m_{ee}^* m_{\mu\mu} \to m_{ee} m_{\mu\mu}^*$.
Thus, for each polarization, the conversion probability~\eqref{eq:prob} is given by
\beq
\begin{aligned}
P(\cM \to \cMbar)_\text{Majorana} 
&= \frac{18432}{(m_\mu a_0)^6} 
   \left| \frac{m_{ee} m_{\mu\mu}}{m_\mu^2} \right|^2 \! 
   \left[ \left( 4 \log\frac{m_W}{m_\mu} + \log\frac{\Lambda}{m_W} \right)^{\!\!2} +4\pi^2 \right]   \\
& = 8.0 \times 10^{-56} 
    \!\left( \frac{| m_{ee} m_{\mu\mu} |^2}{( 10^{-1}\>\mathrm{eV})^2} \right)^{\!\!2} 
    \frac{[ 54.4 + \log(\Lambda / 10^{14}\,\mathrm{GeV}) ]^2 + 4\pi^2}{54.4^2 + 4\pi^2}
\,. \label{eq:pm} \end{aligned}
\eeq
It is interesting to note that the combination~$|m_{ee} m_{\mu\mu}|^2$ also depends on the Majorana phases~\cite{Bilenky:2001rz, Dery:2024lem}.

\subsection{Pseudo--Dirac neutrinos}
\label{sec:prob_pseudodirac}
\begin{figure}[t]
\centering
    \begin{tikzpicture}
        \begin{feynman}
            \vertex (i1) at (-3, 1) {\(\mu^+\)};
            \vertex (i2) at (-3,-1) {\(e^-\)};
            \vertex (o1) at (3, 1) {\(\mu^-\)};
            \vertex (o2) at (3,-1) {\(\e^+\)};

            \vertex (v1) at (-1.5, 1);
            \vertex (v2) at (1.5, 1);
            \vertex (v3) at (-1.5,-1);
            \vertex (v4) at (1.5,-1);
            \vertex (u1) at (-0.75,2);
            \vertex (u2) at (-0.75,-2);
            \vertex (u3) at (0.75,2);
            \vertex (u4) at (0.75,-2);

            \vertex (m11) at (0, 0.5);
            \vertex (m22) at (0, -0.5);

            \vertex [crossed dot](m1) at (0,1) {};
            \vertex [crossed dot](m2) at (0,-1) {};

            \diagram* {
                (i1) -- [anti fermion] (v1) --[anti fermion](m1) -- [fermion] (v2) -- [ fermion] (o1),
                (i2) -- [fermion] (v3) -- [fermion] (m2)--[anti fermion] (v4) -- [anti fermion] (o2),
                (u1) --[dashed] (m1),
                (u2) --[dashed] (m2),
                (u3) --[dashed] (m1),
                (u4) --[dashed] (m2),
            };

            \draw[decorate, decoration={snake, amplitude=0.15cm, segment length=0.35cm}] (v1) -- (v3);
            \draw[decorate, decoration={snake, amplitude=0.15cm, segment length=0.35cm}] (v2) -- (v4);

             \node[] at (m11) {\scriptsize\({\cal L}^*_5\)};
             \node[] at (m22) {\scriptsize\({\cal L}_5\)};
             \node[above] at (u1) {\scriptsize\(\langle H \rangle\)};
             \node[below] at (u2) {\scriptsize\(\langle H \rangle\)};
             \node[above] at (u3) {\scriptsize\(\langle H \rangle\)};
             \node[below] at (u4) {\scriptsize\(\langle H \rangle\)};

        \end{feynman}
    \end{tikzpicture}
\caption{\label{fig:pseudo_dirac} One of the Majorana type
$\cM$--$\cMbar$ mixing diagrams when neutrinos are pseudo-Dirac. ${\cal L}_5$, defined in Eq.~\eqref{eq:L5}, represents the insertion of a small Majorana mass.}
\end{figure} 
Neutrinos are called pseudo--Dirac (or quasi-Dirac) when they have a Majorana mass $m_{\mathrm{M}}$ which is much smaller  than the Dirac masses $m_\mathrm{D}$. 
Theories of pseudo-Dirac neutrinos \cite{Wolfenstein:1981kw, Petcov:1982ya, Valle:1983dk} have 6 light neutrino fields $\nu_{e,\mu,\tau}, N_{1,2,3}$, where $\nu_\alpha$ is the upper component of the lepton doublet $L_\alpha$ and the $N$'s are completely neutral under all SM gauge groups.
With small Majorana mass terms of the form $\nu\nu$, $NN$, or both, 
the spectrum of the theory consists of three quasi-degenerate pairs of neutrinos $\nu_i, \nu_{i'}$, for each $i$. 

For the $\nu$ fields, the small Majorana mass can be imparted, for example, by a dimension-5 operator
\beq
\mathcal{L}_5 = -\frac{Y_{\alpha \beta}}{2\Lambda} (H L_{\alpha}) (H L_{\beta}) \label{eq:L5}
\eeq
where $\alpha$ and $\beta$ are generation indices in the basis where the charged lepton mass matrix is diagonal, 
and $\Lambda$ is roughly the scale of new physics that is integrated out to obtain this operator at low energies. 
The above operator gives rise to a Majorana mass matrix when the Higgs acquires a vev.

The small Majorana mass contributes to $\cM$--$\cMbar$ mixing, as shown in Fig.~\ref{fig:pseudo_dirac}. Ignoring neutrino masses in the propagators, we find that the box functions from Fig.~\ref{fig:pseudo_dirac} is the same as what we found for the Majorana type diagrams in Eq.~\eqref{eq:boxmajorana}, with the replacement $m_{ee}^* m_{\mu \mu} \to Y_{ee} Y_{\mu \mu}^{*} v^4/\Lambda^2$ with $v \approx 174$~GeV\@. Consequently, $\Sigma_{12}$ has the same form as in Eq.~\eqref{eq:Sigma21:Majo}, 
with the replacement $m_{ee}^* m_{\mu\mu} \to \tilde{m}^*_{ee} \tilde m_{\mu \mu}$, where
\beq 
\tilde{m}_{\ell\ell} \equiv \sum_{i,j} U_{\ell i} \, (m_{\mathrm{M}})_{ij} U_{\ell j}
\,. \label{eq:tildemll}
\eeq
Here, the matrix $\mm_{ij}$ is the Majorana mass matrix from ${\cal L}_5$ in the basis where the neutrinos' Dirac mass matrix is diagonal:
\beq {\cal L}_{5, \mathrm{mass}} = -{ U_{\alpha i} Y_{\alpha \beta} U_{\beta j} v^2 \over 2 \Lambda} \nu_i \nu_{j} \equiv -{1 \over 2} (m_{\mathrm{M}})_{ij} \nu_i \nu_{j} \, .
\eeq
The $N_i$ fields do not participate in $\cM$--$\cMbar$ mixing through box diagrams because they do not couple to the $W$. 
Therefore, for the purposes of the present work, we set the Majorana masses of these fields to 0.

While the Majorana mass $m_\mathrm{M}$ is much smaller than the Dirac mass $m_\mathrm{D}$ for pseudo-Dirac neutrinos, the advantage of the diagram in Fig.~\ref{fig:pseudo_dirac} over the Dirac-type diagrams is that it goes as $m_{\mathrm{M}}^2$, while the Dirac-type diagrams go as $(m_\mathrm{D}^4/m_{\mu}^2)$. 
Therefore, as long as $m_{\mathrm{M}}^2 \gg m_\mathrm{D}^2 (m_\mathrm{D}/m_{\mu})^2$, which can easily be satisfied under the pseudo-Dirac condition $m_\mathrm{M} \ll m_\mathrm{D} \sim m_\nu$, the contribution from the Majorana components to $\cM$--$\cMbar$ mixing is larger than in the case where neutrinos are pure Dirac. In such a scenario, the $\cM$--$\cMbar$ mixing probability is given by Eq.~\eqref{eq:pm}, with the replacement $m_{ee}^* m_{\mu \mu} \to \tilde{m}_{ee}^* \tilde{m}_{\mu \mu}$. The question we ask now is how large this can be compared to the usual Dirac case, where we found that the mixing probability is vanishingly small at $\sim 10^{-95}$.

Since the Majorana components in the mass matrix are by definition subdominant to the Dirac components in the pseudo-Dirac case, we can treat the former as perturbation.
To first order in the Majorana components, 
the mass-squared difference $\delta m_{i'i}^2 \equiv m_{i'}^2 - m_i^2$ between states in the $i^\text{th}$ quasi-degenerate pair ($i=1,2,3$) is given by:
\beq 
\delta m_{i'i}^2 = 2 \mm_{ii} m_i \, , 
\eeq
where $m_i$ is the Dirac mass of the $i^\text{th}$ pair.
These mass splittings are constrained by solar and atmospheric neutrino oscillation data~\cite{Cirelli:2004cz}, 
and the most up-to-date bound from the solar neutrino gives $\delta m_{2'2}^2 < 10^{-11}\>\text{eV}^2$~\cite{ansarifard:2022kvy}, 
while the atmospheric neutrino oscillations constrain $\delta m_{3'3}^2 < 10^{-4}\> \text{eV}^2$~\cite{Cirelli:2004cz}. 
To demonstrate the possibility of enhancing $\cM$--$\cMbar$ mixing, we look at the most optimistic scenario where we maximally evade the solar and atmospheric neutrino constraints by setting $\mm_{ii} = 0$ for all $i$. 
In this limit, the mass-squared differences $\delta m_{i'i}^2$ are non-zero beginning only at third order in the off-diagonal elements.
For the most strongly constrained splitting $\delta m_{2'2}^2$, we have
\beq
\delta m_{2'2}^2 
= \frac{4 \mm_{12} \mm_{13} \mm_{23}}{(m_2^2 - m_1^2)(m_2^2 - m_3^2)} m_2^3  
\,. 
\eeq

To be definite let us assume normal ordering $m_1^2 \ll m_2^2 \ll m_3^2$ for the neutrino masses, and that all of the elements of the $\mm$ matrix are roughly the same order of magnitude. Then, for the purpose of making an estimate, we have
\beq 
\delta m_{2'2}^2 \sim 4 \left( { m_{\mathrm{M}}  \over m_3 }\right)^{\!\!3} m_2 m_3  
\,.
\eeq
Then the constraint $\delta m_{2'2}^2 < 10^{-11} \ \text{eV}^2$ translates to
 \beq
\left( { m_{\mathrm{M}}  \over m_3 }\right)^{\!\!3} 
\lesssim 
{10^{-11} \ \text{eV}^2 \over 4m_2 m_3} 
\sim 6 \times 10^{-9}
\label{eq:mmbound}
\eeq
with $m_2 = 0.008\>$eV and $m_3 = 0.05\>$eV\@.
We then have
\beq 
{P(\cM \to \cMbar)_\text{Pseudo-Dirac} \over P(\cM \to \cMbar)_\text{Majorana}}
\sim
\left( { m_{\mathrm{M}}  \over m_3 } \right)^{\!\!4} 
\lesssim 
10^{-11}\, ,
\eeq
From $P(\cM \to \cMbar)_\text{Majorana} \sim 10^{-55}$ in Eq.~\eqref{eq:pm}, we conclude $P(\cM \to \cMbar)_\text{Pseudo-Dirac} \lesssim 10^{-66}$. This can be a 29 order of magnitude improvement over the case when neutrinos are pure Dirac, where the probability is $\sim 10^{-95}$ as in Eq.~\eqref{eq:pd}.
One can repeat the estimation for the case of inverted mass hierarchy for neutrinos,  and find a 21 order of magnitude improvement. 
Therefore, normal ordering provides better prospects than inverted ordering for $\mmbar$ mixing via Pseudo-Dirac neutrinos.

\subsection{Can a non-unitary $U_\mathrm{PMNS}$ enhance $\cM$--$\cMbar$ mixing?}
\label{sec:non-unitary}
We have seen above that $\cM$--$\cMbar$ mixing is extremely suppressed.
The suppression is especially severe for the Dirac type diagrams, where the amplitude is  proportional to $\sim m_\nu^4/m_\mu^2$.
As noted already, this necessity of picking up one factor of $m_\nu^2$ from each of the two neutrino propagators is due to the unitarity of $U_\mathrm{PMNS}$, 
$\sum_i U_{ei}^* U_{\mu i}^{\phantom*} = 0$.  
However, since we do not expect the unitarity of $U_\mathrm{PMNS}$ among the three light neutrinos to be strictly true in the presence of generic new physics, 
we must ask whether a non-unitary $U_\mathrm{PMNS}$ can enhance $\cM$--$\cMbar$ mixing.
Since the unitary part of the contribution is so small, even a slight violation of unitarity might give rise to a dramatic enhancement.
Clearly, such enhancement would completely alter our result for Dirac neutrinos, 
where only the Dirac type diagrams contribute.       

For Majorana neutrinos, the Majorana type diagrams dominate over the Dirac type diagrams with a unitary $U_\mathrm{PMNS}$ as the former only pick up $m_\nu$, not $m_\nu^2$ as in the latter, from each neutrino propagator.
Note that we need to pick up one factor of $m_\nu$ per propagator solely because that is what we mean by the Majorana type, not because of unitarity ($\sum_i U_{e i} U_{e i} \neq 0$ without $m_i$).
Hence, non-unitary $U_\mathrm{PMNS}$ would not enhance the Majorana type diagrams themselves.
Rather, our question is whether non-unitary Dirac type diagrams could be larger than the Majorana type diagrams.
Again, if true, that would completely change our conclusion.

To investigate those important questions, 
consider an effective theory in which we have already integrated out whatever new physics generates non-unitary contributions in $U_\mathrm{PMNS}$.    
In order to have a non-unitary $U_\mathrm{PMNS}$, 
we need a misalignment between the basis in which the left-handed neutrino kinetic term is diagonal 
and that in which the $W$-$\ell_L$-$\nu_L$ coupling is diagonal.
The leading non-renormalizable interaction that can generate such misalignment is found at dimension-6:
\beq
\mathcal{L}_6 = \frac{\lambda_{\alpha\beta}}{M^2} (H^{\herm} \overline{L}_\alpha) \I\slashed{\partial} (H L_\beta)
\,,  \label{eq:L6}
\eeq
where $\alpha, \beta$ are the generation indices in the flavor basis of neutrinos in which the charged lepton mass matrix is diagonal, and $M$ is a proxy for the scale of the new physics we have integrated out, not to be confused with the muonium mass.
The product $HL$ is taken to be completely neutral under the SM gauge group. 
Upon electroweak symmetry breaking, $HL$ contains only a neutrino and no charged lepton in the unitarity gauge.
One could consider variations of $\mathcal{L}_6$ where that is not the case, 
e.g., taking $HL$ to be in an electroweak triplet. 
Those variations would not affect our argument below.
Moreover, they would likely lead to a phenomenological disaster as they would also modify the charged leptons' couplings to the $W$ or $Z$.   
Upon electroweak symmetry breaking with $\langle H^\dag H \rangle = v^2$, 
$\mathcal{L}_6$ modifies the left-handed neutrino kinetic term as
\beq
\delta_{\alpha\beta} \, \overline{\nu}_{\alpha}\I\slashed{\partial} \nu_{\beta}^{\phantom\dag}
\>\longrightarrow\>
\biggl( \delta_{\alpha\beta} + \frac{v^2}{M^2} \lambda_{\alpha\beta} \biggr) 
\overline{\nu}_{\alpha} \I\slashed{\partial} \nu_{\beta}^{\phantom\dag}
\,,
\eeq
which then modifies $U_\mathrm{PMNS}$ as
\beq
U_\mathrm{PMNS} 
\>\longrightarrow\>
U_\mathrm{PMNS} = \biggl( \bm{1} - \frac{v^2}{2M^2} \lambda \biggr) U_\mathrm{PMNS}^{(0)} 
\,, \label{eq:deltaupmns}
\eeq
where $U_\mathrm{PMNS}^{(0)}$ is the unitary PMNS matrix we would have in the absence of $\mathcal{L}_6$, and corrections higher order in $v^2/M^2$ are ignored.
\begin{figure}[t]
\centering
    \hfill
    \begin{subfigure}[b]{0.45\textwidth}
    \begin{tikzpicture}[baseline=(current bounding box.center)]
        \begin{feynman}
            \vertex (i1) at (-3, 1) {\(\mu^+\)};
            \vertex (i2) at (-3,-1) {\(e^-\)};
            \vertex (o1) at (3, 1) {\(e^+\)};
            \vertex (o2) at (3,-1) {\(\mu^-\)};

            \vertex (v1) at (-1.5, 1);
            \vertex (v2) at (1.5, 1);
            \vertex (v3) at (-1.5,-1);
            \vertex (v4) at (1.5,-1);
            \vertex (u1) at (-0.75,2);
            \vertex (u2) at (-0.75,-2);
            \vertex (u3) at (0.75,2);
            \vertex (u4) at (0.75,-2);

            \vertex (m11) at (0, 0.5);
            \vertex (m22) at (0, -0.5);

            \vertex [crossed dot](m1) at (0,1) {};
            \vertex [crossed dot](m2) at (0,-1) {};

            \diagram* {
                (i1) -- [anti fermion] (v1) --[anti fermion](m1) -- [anti fermion] (v2) -- [ anti fermion] (o1),
                (i2) -- [fermion] (v3) -- [fermion] (m2)--[fermion] (v4) -- [fermion] (o2),
                (u1) --[dashed] (m1),
                (u2) --[dashed] (m2),
                (u3) --[dashed] (m1),
                (u4) --[dashed] (m2),
            };

            \draw[decorate, decoration={snake, amplitude=0.15cm, segment length=0.35cm}] (v1) -- (v3);
            \draw[decorate, decoration={snake, amplitude=0.15cm, segment length=0.35cm}] (v2) -- (v4);

             \node[] at (m11) {\scriptsize\({\cal L}_6\)};
             \node[] at (m22) {\scriptsize\({\cal L}_6\)};
             \node[above] at (u1) {\scriptsize\(\langle H \rangle\)};
             \node[below] at (u2) {\scriptsize\(\langle H \rangle\)};
             \node[above] at (u3) {\scriptsize\(\langle H \rangle\)};
             \node[below] at (u4) {\scriptsize\(\langle H \rangle\)};

        \end{feynman}
    \end{tikzpicture}
    \end{subfigure}
     \begin{subfigure}[b]{0.45\textwidth}
    \begin{tikzpicture}[baseline=(current bounding box.center)]
        \begin{feynman}
            \vertex (i1) at (-3, 1) {\(\mu^+\)};
            \vertex (i2) at (-3,-1) {\(e^-\)};
            \vertex (o1) at (3, 1) {\(e^+\)};
            \vertex (o2) at (3,-1) {\(\mu^-\)};

            \vertex (v1) at (-1.5, 0.5);
            \vertex (v2) at (1.5, 0.5);
            \vertex (v3) at (-1.5,-0.5);
            \vertex (v4) at (1.5,-0.5);
            \vertex [crossed dot] (c) at (0,0) {};
            \vertex (c1) at (0,-0.5);

            \diagram* {
                (i1) -- [anti fermion] (c) --[anti fermion](o1),
                (i2) -- [fermion] (c) --[fermion] (o2),
            };


            \node[] at (c1) {\scriptsize\({\cal L}_6'\)};

        \end{feynman}
    \end{tikzpicture}
    \end{subfigure}
\caption{\label{fig:non-unitary}
(Left) One of the Dirac type diagrams with insertions of the ${\cal L}_6$ operator, defined in Eq.~\eqref{eq:L6}, that violate unitarity of $U_\mathrm{PMNS}$. (Right) Any operator such as ${\cal L}_6 $ will also require a dimension-6 operator ${\cal L}_6'$ as given in Eq.~\eqref{eq:L6prime} for consistent renormalization, which will overpower the contribution of the left diagram to $\mmbar$ mixing, completely obscuring the parametric dependence of $\cM$--$\cMbar$ conversion on neutrino properties.}
\end{figure} 
We can now estimate non-unitary contributions to the Dirac type diagrams. 
They come with one factor of $\sum_i U^*_{ei} U_{\mu i}^{\phantom*}$ (now without $m_i^2$) from each neutrino propagator.
Since there is no need for $m_i$, we can simply ignore $m_i$ and work in the flavor basis, 
directly inserting an $\mathcal{L}_6$ into each of the two massless neutrino propagators (See Fig.~\ref{fig:non-unitary} left).
The loop integral is now quadratically divergent in the UV, 
and the box diagrams go as
\beq
\sim \frac{\lambda^2}{16\pi^2} \frac{m_W^2}{M^4} \log\frac{M}{m_W} + \frac{c}{M^2}
\,, \label{eq:quaddivbox}
\eeq
where $c$ is a regulator/scheme dependent constant and hence incalculable, while the coefficient of the logarithm is calculable in the effective theory. 
The $c$ dependence implies the existence of another dimension-6 interaction of the form
\beq
\mathcal{L}'_6 \sim \frac{c'}{M^2} \bar{\mu} e \, \bar{\mu} e \label{eq:L6prime}
\,,
\eeq
for consistent renormalization, where $c'$ is an independent coefficient in the effective theory, completely unrelated to $\lambda_{\alpha\beta}$ or $U_\mathrm{PMNS}$. Of course, in a specific UV-complete model, the coefficients $c$ and $c'$ can be related, but this relationship is model-dependent. As an example, consider a UV completion with a single heavy Dirac singlet $N$ of mass $M_N$,
\beq {\cal L} \supset - M_N \overline{N} N - g_{\alpha} H^{\herm} \overline{L}_\alpha P_R N  
\,. 
\eeq
Once the heavy Dirac singlet is integrated out, this model gives us ${\cal L}_6$ with 
\beq
\lambda_{\alpha \beta} \sim g^*_\alpha g_\beta
\,,\quad M = M_N
\eeq
up to a numerical factor, and also gives
\beq 
c' \sim c \sim \frac{(g_e^* g_\mu)^2}{16 \pi^2} 
\sim \frac{\lambda_{e \mu}^2}{16\pi^2} \,.
\eeq
On the other hand, if we instead consider a different UV model with a heavy weak-triplet scalar $\Phi$,  
\beq
\mathcal{L} \supset -M_\Phi^2 |\Phi|^2 - g_{\alpha\beta} \Phi L_\alpha L_\beta \,,
\eeq
then, we have 
\beq
\lambda_{\alpha\beta} \sim \sum_\gamma \frac{g_{\alpha\gamma} y_\gamma^2 g^*_{\gamma\beta}}{16\pi^2}
\,,\quad
M = M_\Phi
\eeq
where $y_\gamma$ is the charged lepton Yukawa coupling for the lepton $\gamma$ and 
\beq
c' \sim g_{ee} g^*_{\mu\mu} \gg c \sim \frac{\lambda_{e \mu}^2}{16\pi^2} \,.
\eeq
Hence, in each of the above UV models, the $\cM$--$\cMbar$ conversion rate and the unitarity violation in $U_{\rm PMNS}$ are related to UV parameters, 
but the relations are completely different, as expected from the earlier general EFT argument.
Therefore, it is not possible to use $\cM$--$\cMbar$ as a model-independent probe of unitarity violation in $U_{\rm PMNS}$.

To summarize, while in the effective theory $c$ is incalculable and $c'$ is an independent parameter, they may be calculated in specific UV-complete models, and $\cM$--$\cMbar$ can offer a window into the UV physics, but not necessarily into the physics of light neutrinos in which we are interested. Of course, other UV completions that give rise to the operators $\mathcal{L}'_6$ as well as other operators such as $\sim \bar{\mu} e \, \bar{\nu}_{e,\mu} \nu_{e,\mu}$ are interesting probes of new physics in their own right and have been extensively studied in~\cite{Conlin:2020veq, Conlin:2022sga}.
However, these directions go beyond the scope of this paper, i.e., $\cM$--$\cMbar$ as a probe of the low energy properties of neutrinos such as masses, mixing, and Dirac vs Majorana. 

We therefore conclude that in the context of our question---whether $\cM$--$\cMbar$ mixing can probe the low energy properties of neutrinos in a UV model independent way---\emph{it is not possible to enhance $\cM$--$\cMbar$ mixing by non-unitary $U_\mathrm{PMNS}$.}
This conclusion cannot be changed by the introduction of an additional symmetry to forbid $\mathcal{L}'_6$, 
because we need $\mathcal{L}'_6$ for consistent renormalization of the loop using $\mathcal{L}_6$ twice, as we have seen above.

\section{Concluding remarks}
\label{sec:conc}
In the SM minimally extended with massive neutrinos, $\mmbar$ mixing happens via box diagrams with neutrinos in the loop. 
The mixing amplitude depends on neutrino masses, mixing angles, and Dirac-vs-Majorana. 
Thus, if there is no additional new physics that affects the neutrino sector, $\mmbar$ mixing could in principle provide insight into the properties of neutrinos. 

In this work, we calculated the $\mmbar$ mixing amplitude for Dirac, Majorana, and pseudo-Dirac neutrinos. We also studied $\mmbar$ conversion when the lepton mixing matrix is no longer unitary. 
In the case of Majorana neutrinos, we found that our calculation of the $\mmbar$ mixing amplitude differs from the existing literature in the treatment of infrared scales in the problem; the correct IR scale should be the muon mass, and not the neutrino masses.
Taking into account constraints from neutrino oscillation data, we also showed that the rate for pseudo-Dirac neutrinos can be enhanced over the rate for ``pure'' Dirac neutrinos by up to 29 orders of magnitude. 
In contrast, and perhaps most surprisingly, we showed that enhancing the Dirac mixing amplitude by introducing non-unitarity in $U_\mathrm{PMNS}$ is possible only at the cost of masking low energy properties of neutrinos.  

Finally, we pointed out that detecting $\mmbar$ conversion using the kinematics of the decay products of muonium/antimuonium (as in MACS and MACE) could be problematic. 
In an upcoming work \cite{paper-in-preparation} we will examine this background in more detail. 

In summary, $\mmbar$ mixing is an interesting and subtle probe of neutrino properties. 
We hope that this work inspires the community to further explore unconventional observables to study neutrinos.

\acknowledgments
We would like to thank Yuval Grossman for his encouragement in the completion of this work. This work is supported in part by the US Department of Energy grant DE-SC0010102. The work of T.O.~and K.T.~is supported in part by JSPS Grant-in-Aid for Scientific Research (Grant No.\,21H01086).

\appendix

\section{Computation of the bound state $\hat{\Gamma}$ function}
\label{app:bound}
In this appendix, we construct a muonium operator using the fundamental field operators of the SM and show how to calculate the $\hat{\Gamma}$ function used in the main text. 
For a review of bound states and their treatment in field theory, see for example \cite{Salpeter:1951sz,Weinberg_1995,Duncan:2012aja,Hoyer:2014gna,Hoyer:2016aew}.

Operators of the form $\mathcal{O} = \overline{\mu} V e$, for various $V$, will have nonzero overlap between muonium states and the vacuum.
For example, with $V = \gamma^\mu$ the operator will have overlap with spin-1 muonium (ortho-muonium), while $V=\gamma_5$ will have overlap with spin-0 muonium (para-muonium).
Since the ground-state muonium can take either spin, from now on we will take $\mathcal{O}^\mu \equiv \bar{\mu} V^\mu e$ with $V^\mu \equiv \gamma^{\mu} P_L$ so that we can access both spin states simultaneously. 
For example, the scalar muonium operator $\phi$ used in Section~\ref{sec:review} can be expressed in terms of the operator $\mathcal{O}^{\mu}$ as
\begin{equation}
    \begin{aligned}
        \phi = \frac{1}{M} \partial_{\mu} \!\left( \frac{\mathcal{O}^{\mu}}{\sqrt{Z}} \right),
    \end{aligned}
    \label{eq:scalar-op}
\end{equation}
where $\sqrt{Z}$ is the amplitude for $\mathcal{O}^{\mu \dagger}$ to create muonium from the vacuum. Calculating $Z$, which is often referred to as the wave function renormalization, will be important in the determination of the mixing rate. One can calculate the wave function renormalization for an operator by examining the two-point correlation function. In momentum space, the two-point function has a pole at every $P^2 = M^2$ corresponding to a physical particle state. The K{\"a}ll{\'e}n-Lehmann spectral representation of the two-point function shows that the residue at such a pole is equal to $Z$. Thus, our first goal is to calculate the two-point correlation function of $\mathcal{O}^{\mu}$, which we refer to as $\mathbf{\Delta}^{\mu \nu}$, and to extract $Z$ as the residue at the muonium bound state pole. The two-point function is
\begin{equation}
    \begin{aligned}
        \mathbf{\Delta}^{\mu \nu}(P^2) = \int\! d^4x \braket{\Omega|T\{ \mathcal{O}^{\mu}(x) \,\mathcal{O}^{\nu \dagger}(0)\}|\Omega} e^{iP \cdot x}.
    \label{eq:two-point}
    \end{aligned}
\end{equation}
The two-point function can be calculated diagrammatically as shown in Fig. \ref{fig:Delta}. 

\begin{figure}[t]
    \centering
    \begin{tikzpicture}[thick, scale=1]

        \node at (0,0) {$\mathbf{\Delta}^{\mu \nu}(P^2)$};
        \node at (1, 0) {$ = $};
        
        \draw[<-] (1.5, 0) -- (2, 0);
        \node at (1.85, 0.3) {\small $V^{\mu}$};
        \node at (1.75, -0.25) {$P$};
        
        \begin{feynman}[scale=1]
            \vertex (in) at (2, 0);
            \fill (in) circle (2pt);
            \vertex (out) at (3, 0);
            \fill (out) circle (2pt);

            \diagram* {
                (in) -- [fermion, dashed, half left] (out),
                (in) -- [anti fermion, half right] (out),
            };
        \end{feynman}

        \draw[<-] (3, 0) -- (3.5, 0);
        \node at (3.25, 0.3) {\small $V^{\nu}$};
        \node at (3.25, -0.25) {$P$};

        \node at (4, 0) {$ + $};

        \draw[<-] (4.5, 0) -- (5, 0);
        \node at (4.80, 0.3) {\small $V^{\mu}$};
        \node at (4.75, -0.25) {$P$};

        \begin{feynman}[scale=1]
            \vertex (in) at (5, 0);
            \fill (in) circle (2pt);
            \vertex (p1) at (6, 1);
            \vertex (p2) at (6, -1);
            \vertex (out) at (7, 0);
            \fill (out) circle (2pt);

            \diagram* {
                (in) -- [fermion, dashed, quarter left] (p1) -- [fermion, dashed, quarter left] (out),
                (in) -- [anti fermion, quarter right] (p2) -- [anti fermion, quarter right] (out),
                (p1) -- [photon] (p2),
            };
        \end{feynman}

        \draw[<-] (7, 0) -- (7.5, 0);
        \node at (7.25, 0.3) {\small $V^{\nu}$};
        \node at (7.25, -0.25) {$P$};

        \node at (8, 0) {$ + $};

        \draw[<-] (8.5, 0) -- (9, 0);
        \node at (8.80, 0.3) {\small $V^{\mu}$};
        \node at (8.75, -0.25) {$P$};

        \begin{feynman}[scale=1]
            \vertex (in) at (9, 0);
            \fill (in) circle (2pt);
            \vertex (p1) at (10, 1);
            \vertex (p2) at (10, -1);
            \vertex (p3) at (11, 1);
            \vertex (p4) at (11, -1);
            \vertex (out) at (12, 0);
            \fill (out) circle (2pt);

            \diagram* {
                (in) -- [fermion, dashed, quarter left] (p1) -- [fermion, dashed] (p3) -- [fermion, dashed, quarter left] (out),
                (in) -- [anti fermion, quarter right] (p2) -- [anti fermion] (p4) -- [anti fermion, quarter right] (out),
                (p1) -- [photon] (p2),
                (p3) -- [photon] (p4),
            };
        \end{feynman}

        \draw[<-] (12, 0) -- (12.5, 0);
        \node at (12.25, 0.3) {\small $V^{\nu}$};
        \node at (12.25, -0.25) {$P$};

        \node at (13, 0) {$ + \cdots$};
        
\end{tikzpicture}
    \caption{Muonium two-point function under the ladder approximation. $P$ is the muonium center of mass 4-momentum. In this approximation, all diagrams with uncrossed photon lines are summed. Dashed lines represent the muon propagators, while solid lines are the electron propagators.}
    \label{fig:Delta}
\end{figure}

Several important comments are in order:

\begin{enumerate}
    \item None of the diagrams which contributes to $\Delta^{\mu \nu}$ have a pole at $P^2 = M^2$. Since a pole cannot arise from summing a finite number of diagrams, none of which contain the pole, we will necessarily need to sum an infinte set of diagrams to see the muonium bound state pole emerge.
    \item The electrons and muons in this picture are off-shell due to the binding energy of muonium. 
    \item  When calculating the two-point function to lowest order in $\alpha$, we need to consider \textit{all} of the ladder diagrams in which photon lines do not cross. Since the relative 3-momentum between the electron and antimuon is of the order of the inverse Bohr radius, we have $v \sim \alpha$ for the relative velocity $v$ between $\mu^+$ and $e^-$. Then, to lowest order in $\alpha$, all propagators go as $v^{-2} \sim \alpha^{-2}$, and every loop comes with a $d^4 q = dq^0 d^3 \mathbf{q} \sim v^2 v^3 \sim \alpha^{5}$. The photon-fermion vertices also contribute an $\alpha$ for every photon line in the diagram. The total number of propagators in diagrams with $n$ uncrossed photon lines is $3n + 2$ while the total number of loops is $n + 1$. Thus, the order in $\alpha$ of a diagram with $n$ uncrossed photon lines is given by $-2(3n + 2) + 5(n + 1) + n = 1$, so all diagram consisting of uncrossed photon exchanges are of order $\alpha$. Since all of these diagram are the same order in the coupling, all must be included when calculating the two-point function. This gives an infinite set of diagrams to sum, which, as mentioned above, is a necessary condition to see the muonium pole emerge. 
    \item Because of the analytic structure, diagrams with crossed photon lines can be ignored to lowest order in $\alpha$.  In the non-relativistic (NR) approximation, each fermion propagator only has one pole. In the case of uncrossed photon lines, the poles from the electron and muon propagators lie on opposite sides of the real axis in the complex energy plane. Thus, we are forced to pick up one of the poles when closing the contour to integrate out the energy component of the loop momentum. In the case of crossed photon lines, the electron and muon poles are on the same side of the real axis. Hence, we can close the contour in the empty half plane, avoiding both poles. In short, the crossed photon diagrams do not contribute at leading order in $\alpha$.
    \item We do not include QED corrections such as the vacuum polarization, fermion self-energy, or vertex corrections, which are higher order in $\alpha$. 
    \item We also do not include the muon decay width in the muon propagator since the width is much smaller than the muonium binding energy.
\end{enumerate}

\begin{figure}[t]
    \centering
    \begin{tikzpicture}[thick, scale=1]

        \node at (0,0) {$\Gamma^{\nu}(P,k)$};
        \node at (1, 0) {$ = $};
        
        \begin{feynman}[scale=1]
            \vertex (in) at (3, 0);
            \fill (in) circle (2pt);
            \vertex (out_top) at (1.5, 1);
            \vertex (out_bottom) at (1.5, -1);            

            \diagram* {
                (in) -- [anti fermion, dashed, quarter right, looseness=1.5] (out_top),
                (in) -- [fermion, quarter left, looseness=1.5] (out_bottom),
            };
        \end{feynman}

        \draw[<-] (3.1, 0) -- (3.6, 0);
        \node at (3.25, 0.3) {\small $V^{\nu}$};
        \node at (3.35, -0.25) {$P$};

        \node at (4, 0) {$ + $};

        \begin{feynman}[scale=1]
            \vertex (in) at (7, 0);
            \fill (in) circle (2pt);
            \vertex (p1) at (6, 1);
            \vertex (p2) at (6, -1);
            \vertex (out_top) at (5, 1);
            \vertex (out_bottom) at (5, -1);

            \diagram* {
                (in) -- [anti fermion, dashed, quarter right] (p1) -- [anti fermion, dashed] (out_top),
                (in) -- [fermion, quarter left] (p2) -- [fermion] (out_bottom),
                (p1) -- [photon, reversed momentum'=\(q\)] (p2),
            };
        \end{feynman}

        \draw[<-] (7.1, 0) -- (7.6, 0);
        \node at (7.25, 0.3) {\small $V^{\nu}$};
        \node at (7.35, -0.25) {$P$};

        \node at (8, 0) {$ + $};

        \begin{feynman}[scale=1]
            \vertex (in) at (12, 0);
            \fill (in) circle (2pt);
            \vertex (p1) at (11, 1);
            \vertex (p2) at (11, -1);
            \vertex (p3) at (10, 1);
            \vertex (p4) at (10, -1);
            \vertex (out_top) at (9, 1);
            \vertex (out_bottom) at (9, -1);
            \fill (out) circle (2pt);

            \diagram* {
                (in) -- [anti fermion, dashed, quarter right] (p1) -- [anti fermion, dashed] (p3) -- [anti fermion, dashed] (out_top),
                (in) -- [fermion, quarter left] (p2) -- [fermion] (p4) -- [fermion] (out_bottom),
                (p1) -- [photon] (p2),
                (p3) -- [photon] (p4),
            };
        \end{feynman}

        \draw[<-] (12.1, 0) -- (12.6, 0);
        \node at (12.25, 0.3) {\small $V^{\nu}$};
        \node at (12.35, -0.25) {$P$};

        \node at (13.25, 0) {$ + \cdots$};


        \node at (1, -3) {$ = $};
        
        \begin{feynman}[scale=1]
            \vertex (in) at (3, -3);
            \fill (in) circle (2pt);
            \vertex (out_top) at (1.5, -2);
            \vertex (out_bottom) at (1.5, -4);            

            \diagram* {
                (in) -- [anti fermion, dashed, quarter right, looseness=1.5] (out_top),
                (in) -- [fermion, quarter left, looseness=1.5] (out_bottom),
            };
        \end{feynman}

        \draw[<-] (3.1, -3) -- (3.6, -3);
        \node at (3.25, -2.7) {\small $V^{\nu}$};
        \node at (3.35, -3.25) {$P$};

        \node at (4, -3) {$ + $};
        
        \begin{feynman}[scale=1]
            \vertex [draw, circle, minimum size=1cm, fill=gray!30] (in) at (7, -3) {\(\Gamma^{\nu}\)};
            \vertex (p1) at (6, -2);
            \vertex (p2) at (6, -4);
            \vertex (out_top) at (5, -2);
            \vertex (out_bottom) at (5, -4);

            \diagram* {
                (in) -- [anti fermion, dashed, quarter right, looseness=0.5] (p1) -- [anti fermion, dashed] (out_top),
                (in) -- [fermion, quarter left, looseness=0.5] (p2) -- [fermion] (out_bottom),
                (p1) -- [photon, reversed momentum'=\(q\)] (p2),
            };
        \end{feynman}

        \draw[<-] (7.6, -3) -- (8.1, -3);
        \node at (7.85, -3.25) {$P$};
        
\end{tikzpicture}
    \caption{Diagrammatic equation for $\Gamma^{\nu}(P,k)$. The propagators are not included on the external legs. $P$ is the muonium center of mass 4-momentum while $k$ is the relative 4-momentum between the electron and antimuon. Dashed lines represent the muon propagators, while solid lines are the electron propagators.}
    \label{fig:Gamma}
\end{figure}

The calculation of $\mathbf{\Delta}^{\mu \nu}(P^2)$ proceeds by first defining the function $\Gamma^{\nu}(P,k)$ as in Fig.~\ref{fig:Gamma}. This $\Gamma$ module interpolates the muonium field to the standard model $e$ and $\mu$ fields. To calculate $\Gamma^{\nu}(P,k)$, we translate the diagrammatic equation of Fig.~\ref{fig:Gamma} into an algebraic equation by noting that the dashed lines represent the muon field and the solid lines represent the electron field. Note that quantities related to the muons will have a subscript ``$1$" while electron quantities will be denoted with a subscript ``$2$". We work in relative coordinates where the electron momentum is $p_2 = \beta P + k$ and the antimuon momentum is $p_1 = \alpha P - k$ with $P$ the total muonium momentum, $\alpha = m_{\mu} / (m_{\mu} + m_{e})$, and $\beta = 1- \alpha$. We find
\begin{equation}\label{Eq: Gamma Master}
    \Gamma^{\nu}(P,k) = V^{\nu} + \int\! \frac{\D^4  q}{(2\pi)^4} \frac{-i(-ie\gamma^{\mu}) S_{2}(p_2 + q) \Gamma^{\nu}(P, k + q) S_{1}(-p_1 + q) (-ie\gamma_{\mu})}{q^2 + i \epsilon} \, .
\end{equation}

Now since muonium is an electromagnetic bound state, and the relative velocity between the constituents is $v \sim \alpha \ll 1$. 
We will use NR approximations and calculate to lowest order in $v$. 
In the NR limit, $p_1$ and $p_2$ are close to the mass shell, so we introduce $\omega_1$ and $\omega_2$ to parameterize the difference, where $\omega_1 \sim \omega_2 \sim v^2$. Working in the rest frame of the muonium, we can now expand the propagators in the limit $q^0 \sim v^2$:
\begin{equation}\label{Eq: Propagator Approx}
    \begin{aligned}
        S_1(-p_1 + q) &=\frac{-i P^{-}}{q^0 - \omega_1 + \frac{(\mathbf{k} + \mathbf{q})^2}{2 m_1} - i \epsilon} \, ,\\
        S_2(p_2 + q) &= \frac{i P^{+}}{q^0 + \omega_2 - \frac{(\mathbf{k} + \mathbf{q})^2}{2 m_2} + i \epsilon} \, ,
    \end{aligned}
\end{equation}
with $P^{\pm} = (\mathbf{1} \pm \gamma^0)/2$. 
We combine Eq.~\eqref{Eq: Gamma Master} with Eq.~\eqref{Eq: Propagator Approx}, working to lowest order in $v \sim \alpha$. Then we integrate over the $q^0$ component by closing the contour up and picking up the pole at $q^0 = \omega_1 - \frac{(\mathbf{k} + \mathbf{q})^2}{2 m_1}$. This gives us
\begin{equation}\label{Eq: Master Gamma Residue}
    \begin{aligned}
        \Gamma^{\nu}(P,k) &= V^{\nu} + e^2 \int\! \frac{\D^3 \mathbf{q}}{(2\pi)^3} \frac{\gamma^{\mu} P^{+} \Gamma^{\nu}(P, k + q) P^{-} \gamma_{\mu}}{[\omega - \frac{(\mathbf{k} + \mathbf{q})^2}{2 \mu} + i \epsilon][\mathbf{q}^2 - i \epsilon]}\, ,
    \end{aligned}
\end{equation}
where $\omega = \omega_1 + \omega_2$ and $\mu$ is the reduced mass. 
We note that the $q$ in $\Gamma^{\nu}(P,k+q)$ is a only a function of $\mathbf{q}$. 
Now, observe that this $\Gamma$ module cannot stand in isolation, and will be connected to other diagrams of interest by $S_2 \Gamma^{\nu} S_1$. In the NR limit of interest here, $S_1(-\alpha P + k) \propto P^-$ and $S_2(\beta P + k) \propto P^+$. So the quantity we actually need to compute is $P^+ \Gamma^{\nu} P^-$. Sandwiching Eq.~\eqref{Eq: Master Gamma Residue} between $P^+$ and $P^-$ and simplifying the Dirac algebra gives
\begin{equation}\label{Eq: Master Gamma Proj Calc}
    \begin{aligned}
        P^+ \Gamma^{\nu}(P,k) P^- 
        &= P^+ V^{\nu} P^- - e^2 \int\! \frac{\D^3 \mathbf{q}}{(2\pi)^3} \frac{P^+ \Gamma^{\nu}(P, k + q) P^-}{[\omega - \frac{(\mathbf{k} + \mathbf{q})^2}{2 \mu} + i \epsilon][\mathbf{q}^2 - i \epsilon]}
    \end{aligned} \, .
\end{equation}
To proceed, we define $G^{\nu}(P,k)$ by 
\begin{equation}\label{Eq:G Def}
    P^+ \Gamma^{\nu}(P, k) P^- = \left(\omega - \frac{\mathbf{k}^2}{2 \mu} +i\epsilon \right) G^{\nu}(P, k) \, .
\end{equation}
Then in terms of $G^{\nu}$, Eq.~\eqref{Eq: Master Gamma Proj Calc} becomes
\begin{equation}\label{Eq: Master G}
    \begin{aligned}
        \left(\omega - \frac{\mathbf{k}^2}{2 \mu} +i\epsilon \right) G^{\nu}(P,k) + e^2 \int\! \frac{\D^3 \mathbf{q}}{(2\pi)^3} \frac{G^{\nu}(P, k + q)}{\mathbf{q}^2 - i \epsilon} &= P^+ V^{\nu} P^- \,.
    \end{aligned}
\end{equation}
\begin{figure}[t]
    \centering
    \begin{tikzpicture}[thick, scale=1]

        \node at (0,0) {$\mathbf{\Delta}^{\mu \nu}(P^2)$};
        \node at (1, 0) {$ = $};
        
        \draw[<-] (1.4, 0) -- (1.9, 0);
        \node at (1.85, 0.3) {\small $V^{\mu}$};
        \node at (1.65, -0.25) {$P$};
        
        \begin{feynman}[scale=1]
            \vertex [draw, circle, minimum size=1cm, fill=gray!30] (in) at (4, 0) {\(\Gamma^{\nu}\)};
            \vertex (out) at (2, 0);
            \fill (out) circle (2pt);

            \diagram* {
                (in) -- [anti fermion, dashed, quarter right] (out),
                (in) -- [fermion, quarter left] (out),
            };
        \end{feynman}

        \draw[<-] (4.55, 0) -- (5.05, 0);
        \node at (4.8, -.25) {$P$};
        
\end{tikzpicture}
    \caption{Muonium two-point function where the ladder diagrams have been re-summed in the $\Gamma$ module. Dashed lines represent the muon propagators, while solid lines are the electron propagators.}
    \label{fig:Delta2}
\end{figure}
Eq.~\eqref{Eq: Master G} is nothing but the Schr\"odinger equation in $\mathbf{k}$ with a source term, $P^+ V^{\nu} P^-$. The general solution is obtained by convolving the source with the Green's function for the Schr\"odinger operator. Thus we find
\begin{equation}\label{Eq: G Soln}
    G^{\nu}(P,k) = \int \frac{\D^3 \mathbf{k}'}{(2\pi)^3} \sum_{n} \frac{\Tilde{\psi}^{*}_n(\mathbf{k}) \Tilde{\psi}_n(\mathbf{k}')}{\omega - E_n +i\epsilon} P^+ V^{\nu} P^- = \sum_{n} \frac{\Tilde{\psi}^{*}_n(\mathbf{k}) \psi_n(0)}{\omega - E_n +i\epsilon} P^+ V^{\nu} P^- \, ,
\end{equation}
and
\begin{equation}\label{Eq: Gamma Soln}
    P^+ \Gamma^{\nu}(P,k) P^- = \left(\omega - \frac{\mathbf{k}^2}{2 \mu} +i\epsilon \right) \sum_{n} \frac{\tilde{\psi}^{*}_n(\mathbf{k}) \psi_n(0)}{\omega - E_n +i\epsilon} P^+ V^{\nu} P^- \, .
\end{equation}
Here $\psi_n(\mathbf{r})$ is the muonium wave function for the $\ket{n,\ell, m} = \ket{n,0,0}$ state of muonium, and $\tilde{\psi}_n(\mathbf{k})$ its Fourier transform.
Note that we only need these $s$ states since all higher orbital angular momentum wave functions have $\psi(0) = 0$. We can now multiply Eq.~\eqref{Eq: Gamma Soln} by the propagators necessary to connect $\Gamma^{\nu}$ to another module. We make the $k^0$ dependence explicit by writing $\omega_1$ and $\omega_2$ in terms of $k^0$ and $\omega$. We then have
\begin{equation}\label{Eq: Gamma Master NR}
    \begin{aligned}
        S_{2}(p_2) \Gamma^{\mu}(P,k) S_{1}(-p_1) &= \sum_{n} \frac{P^{+} V^{\mu} P^{-}}{\omega - E_n +i\epsilon} \frac{(\omega - \frac{\mathbf{k}^2}{2 \mu}) \Tilde{\psi}^{*}_n(\mathbf{k}) \psi_n(0)}{[k^0 + \beta \omega - \frac{\mathbf{k}^2}{2 m_2} + i \epsilon][k^0 -\alpha \omega + \frac{\mathbf{k}^2}{2 m_1} - i \epsilon]}\, .
    \end{aligned}   
\end{equation}
Using Eq.~\eqref{Eq: Gamma Master NR} we can calculate the muonium two-point function~(\ref{eq:two-point}) by closing the other ends of the propagators with $V^{\mu}$ as shown in Fig.~\ref{fig:Delta2}. Closing in this way results in a trace and a negative sign from the closed fermion loop. After performing the $k$ integration, we find
\begin{equation}\label{Eq: Delta Master NR}
    \begin{aligned}
        \mathbf{\Delta}^{\mu \nu}(P^2) &= -\int \frac{\D^4k}{(2\pi)^4} \Tr\{V^{\mu}S_{2}(p_2) \Gamma^{\nu}(P,k) S_{1}(-p_1)\} \\
        &= -i \sum_{n} \frac{\Tr\{V^{\mu} P^{+} V^{\nu} P^{-}\} \left| \psi_n(0) \right|^2}{\omega - E_n +i\epsilon}
    \end{aligned}   
\end{equation}
Recalling $V^{\mu} = \gamma^{\mu} P_L$ and that we are working in the muonium rest frame, we can evaluate the trace and use
\begin{equation} \label{eq:Non-Rel_Propagator}
    \begin{aligned}
        \frac{1}{\omega - E_n +i\epsilon} = \frac{2 M_n}{P^2 - M_n^2 +i\epsilon} + {\cal O}(v^2)       
    \end{aligned}
\end{equation}
to obtain the following expression for the muonium two-point function:
\begin{equation}\label{eq:Delta_Master}
    \begin{aligned}
        \mathbf{\Delta}^{\mu \nu}(P^2) &= \sum_{n} M_n \left| \psi_n(0) \right|^2 \Biggl[ \frac{i \left( -g^{\mu \nu} + \frac{P^{\mu} P^{\nu}}{M_n^2}\right)}{P^2 - M_n^2 + i\epsilon} + \frac{P^\mu}{M_n} \frac{i}{P^2 - M_n^2 +i\epsilon}\frac{P^\nu}{M_n} \Biggr] \, .
    \end{aligned}   
\end{equation}
The first term inside the square brackets in Eq.~\eqref{eq:Delta_Master} is the propagator for vector muonium. The second term represents the propagator for scalar muonium created by $\phi^\dag$ in Eq.~\eqref{eq:scalar-op}. This confirms that the operator $\mathcal{O}^{\mu \dagger} = \overline{e} \gamma^{\mu} P_L \mu$ creates both scalar and vector muonium. The wave function renormalization for both scalar and vector muonium is $Z_n = M_n |\psi_n(0)|^2$, indicating that both states are created with equal probability.

Now, to make use of $\Gamma^{\mu}$ in our calculation of the $\mmbar$ amplitude we need to determine how $\Gamma^{\mu}$ will interact with the box diagrams. In particular, we note that when attaching to box diagrams, the $S_2$ and $S_1$ propagators always terminate on a weak vertex. Thus, we need to calculate $P_L S_2 \Gamma^{\mu} S_1 P_R$. In the rest frame of muonium, we have $P^{\pm} = (1 \pm \slashed{P}/M_n)/2$. Then, the spinor structure is given by
\begin{equation}\label{Eq: Spin-Structure}
    \begin{aligned}
        P_L P^{+} \gamma^{\mu} P_L P^{-} P_R  &= \frac{1}{4} P_L \left(1 + \frac{\slashed{P}}{M_n}\right) \gamma^{\mu} P_L \left(1 - \frac{\slashed{P}}{M_n}\right) P_R \\
        &= -\frac{1}{4} \frac{P_L \slashed{P} P_R}{M_n} \gamma^{\mu} \frac{P_L \slashed{P} P_R}{M_n} \\
        &= -\frac{P \!\cdot\! \sigma \, \bar{\sigma}^{\mu} P \!\cdot\! \sigma}{4 M_n^2} \otimes 
        \begin{pmatrix}
            0 & 1 \\
            0 & 0
        \end{pmatrix} \\
    \end{aligned}   
\end{equation}
Therefore, we find,
\begin{equation}\label{Eq: Gamma Module (2-Component) Covariant}
    \begin{aligned}
        (P_L S_{2} \Gamma^{\mu}(P,k) S_{1} P_R)_{b \dot{a}} &= -\sum_{n} \frac{(P \!\cdot\! \sigma \, \bar{\sigma}^{\mu} P \!\cdot\! \sigma)_{b \dot{a}} (\omega - \frac{\mathbf{k}^2}{2 \mu}) \Tilde{\psi}^{*}_n(\mathbf{k}) \psi_n(0)}{4 M_n^2 [\omega - E_n + i\epsilon][k^0 + \beta \omega - \frac{\mathbf{k}^2}{2 m_2} + i \epsilon][k^0 -\alpha \omega + \frac{\mathbf{k}^2}{2 m_1} - i \epsilon]} \\
        &= -\sum_{n} \frac{(P \!\cdot\! \sigma \, \bar{\sigma}^{\mu} P \!\cdot\! \sigma)_{b \dot{a}} (\omega - \frac{\mathbf{k}^2}{2 \mu}) \Tilde{\psi}^{*}_n(\mathbf{k}) \psi_n(0)}{2 M_n [P^2 - M_n^2 + i\epsilon][k^0 + \beta \omega - \frac{\mathbf{k}^2}{2 m_2} + i \epsilon][k^0 -\alpha \omega + \frac{\mathbf{k}^2}{2 m_1} - i \epsilon]} \, ,
    \end{aligned}   
\end{equation}
where once again we have made use of \eqref{eq:Non-Rel_Propagator}. Now when the $\Gamma^{\mu}$ module is connected to a box diagram, we have to integrate over the relative momentum $k$ which is now a loop momentum. However, since to lowest order in $v$, the box diagrams are independent of $k$, the box function can be factored out of the integral. Thus, the integral over $k$ of Eq.~\eqref{Eq: Gamma Module (2-Component) Covariant} can be performed immediately:
\begin{equation}\label{Eq: Gamma Module (2-Component) Covariant Integrated}
    \begin{aligned}
        \int\! \frac{\D^4 k}{(2 \pi)^4}(P_L S_{2} \Gamma^{\mu}(P,k) S_{1} P_R)_{b \dot{a}} &= -i \sum_{n} \frac{(P \!\cdot\! \sigma \, \bar{\sigma}^{\mu} P \!\cdot\! \sigma)_{b \dot{a}} \left| \psi_n(0) \right|^2}{2 M_n [P^2 - M_n^2 +i\epsilon]} \, .
    \end{aligned}   
\end{equation}

Finally, we will define $\hat{\Gamma}^{\mu}(P)$ as the result of applying the LSZ reduction to Eq.~\eqref{Eq: Gamma Module (2-Component) Covariant Integrated} in order to amputate the external muonium propagator. 
This procedure puts the muonium on-shell by multiplying with $-i(P^2 - M^2)$  and taking the limit $P^2 \rightarrow M^2$ to remove the pole. 
Here, $M=M_1$ is the mass of the $1s$ state of muonium. 
The LSZ reduction also corrects for the non-unit residue by dividing by $\sqrt{Z}$ corresponding to the pole at $P^2 = M^2$. This finally leaves us with
\begin{equation}\label{Eq: Gamma hat}
    \begin{aligned}
        \hat{\Gamma}_{b \dot{a}}^{\mu}(P) &= \frac{1}{i\sqrt{Z}} \lim_{P^2 \rightarrow M^2} (P^2 - M^2) \int\! \frac{\D ^4 k}{(2 \pi)^4}(P_L S_{2} \Gamma^{\mu}(P,k) S_{1} P_R)_{b \dot{a}} \\
        &= - \frac{\left| \psi(0) \right|}{2 M^{3/2}} (P \!\cdot\! \sigma \, \bar{\sigma}^{\mu} P \!\cdot\! \sigma)_{b \dot{a}} \, .
    \end{aligned}   
\end{equation}
This completes the derivation of Eq.~\eqref{eq:gamma_nu}.


\bibliography{mybib}{}
\bibliographystyle{JHEP}

\end{document}